\newif\iffigs\figsfalse              
\figstrue                            
\newif\ifbbB\bbBfalse                
\bbBtrue                             

\input harvmac
\def\newdate{May 2005}

\newcount\figno \figno=0
\iffigs
 \message{If you do not have epsf.tex to include figures,}
 \message{change the option at the top of the tex file.}
 \input epsf
 \def\fig#1#2#3{\par\begingroup\parindent=0pt\leftskip=1cm\rightskip=1cm
  \parindent=0pt \baselineskip=11pt \global\advance\figno by 1 \midinsert
  \epsfxsize=#3 \centerline{\epsfbox{#2}} \vskip 12pt
  {\bf Fig. \the\figno:} #1\par \endinsert\endgroup\par }
\else
 \def\fig#1#2#3{\global\advance\figno by 1 \vskip .25in
  \centerline{\bf Figure \the\figno} \vskip .25in}
\fi

\def\a{\alpha}\def\b{\beta} \def\g{\gamma} \def\l{\lambda}
\def\d{\delta} \def\e{\epsilon} 
\def\r{\rho}
\def\s{\sigma}   
  
\def\half{{1\over 2}}

\def\id {{\bf 1}}

\def\m{\mu}
\def\n{\nu}

\def\barpsi{\bar{\psi}}

\def\barchi{\bar{\chi}}

\def\alphadot{\dot{\alpha}}
\def\betadot{\dot{\beta}}
\def\gammadot{\dot{\gamma}}
\def\deltadot{\dot{\delta}}

\def\psiu{\psi^{\alpha}}
\def\psiudot{\barpsi^{\alphadot}}
\def\psid{\psi_{\alpha}}
\def\psiddot{\barpsi_{\alphadot}}
\def\epsu{\epsilon^{\alpha\beta}}
\def\epsudot{\epsilon^{\alphadot\betadot}}
\def\epsd{\epsilon_{\alpha\beta}}
\def\epsddot{\epsilon_{\alphadot\betadot}}

\def\leaderfill{\leaders\hbox to 1em{\hss.\hss}\hfill}


\lref\ArakiSE{
T.~Araki, K.~Ito and A.~Ohtsuka,
``Supersymmetric gauge theories on noncommutative superspace,''
Phys.\ Lett.\ B {\bf 573}, 209 (2003)
[arXiv:hep-th/0307076].
}

\lref\SeibergYZ{
N.~Seiberg,
``Noncommutative superspace, N = 1/2 supersymmetry, field theory and  string
theory,''
JHEP {\bf 0306}, 010 (2003)
[arXiv:hep-th/0305248].
}

\lref\GrassiQK{
P.~A.~Grassi, R.~Ricci and D.~Robles-Llana,
``Instanton calculations for N = 1/2 super Yang-Mills theory,''
JHEP {\bf 0407}, 065 (2004)
[arXiv:hep-th/0311155].
}

\lref\ImaanpurJJ{
A.~Imaanpur,
``On instantons and zero modes of N = 1/2 SYM theory,''
JHEP {\bf 0309}, 077 (2003)
[arXiv:hep-th/0308171].
}

\lref\ImaanpurIG{
A.~Imaanpur,
``Comments on gluino condensates in N = 1/2 SYM theory,''
JHEP {\bf 0312}, 009 (2003)
[arXiv:hep-th/0311137].
}

\lref\BrittoUV{
R.~Britto, B.~Feng, O.~Lunin and S.~J.~Rey,
``U(N) instantons on N = 1/2 superspace: Exact solution and geometry of moduli
space,''
Phys.\ Rev.\ D {\bf 69}, 126004 (2004)
[arXiv:hep-th/0311275].
}

\lref\ShifmanMV{
M.~A.~Shifman and A.~I.~Vainshtein,
``Instantons versus supersymmetry: Fifteen years later,''
arXiv:hep-th/9902018.
}

\lref\DoreyIK{
N.~Dorey, T.~J.~Hollowood, V.~V.~Khoze and M.~P.~Mattis,
``The calculus of many instantons,''
Phys.\ Rept.\  {\bf 371}, 231 (2002)
[arXiv:hep-th/0206063].
}

\lref\BelitskyWS{
A.~V.~Belitsky, S.~Vandoren and P.~van Nieuwenhuizen,
``Yang-Mills and D-instantons,''
Class.\ Quant.\ Grav.\  {\bf 17}, 3521 (2000)
[arXiv:hep-th/0004186].
}

\lref\BerkovitsKJ{
N.~Berkovits and N.~Seiberg,
``Superstrings in graviphoton background and N = 1/2 + 3/2 supersymmetry,''
JHEP {\bf 0307}, 010 (2003)
[arXiv:hep-th/0306226].
}

\lref\AffleckMP{
I.~Affleck,
``On Constrained Instantons,''
Nucl.\ Phys.\ B {\bf 191}, 429 (1981);
M.~Nielsen and N.~K.~Nielsen,
``Explicit construction of constrained instantons,''
arXiv:hep-th/9912006.
}

\lref\OoguriQP{
  J.~de Boer, P.~A.~Grassi and P.~van Nieuwenhuizen,
  ``Non-commutative superspace from string theory,''
  Phys.\ Lett.\ B {\bf 574}, 98 (2003)
  [arXiv:hep-th/0302078];
H.~Ooguri and C.~Vafa,
``The C-deformation of gluino and non-planar diagrams,''
Adv.\ Theor.\ Math.\ Phys.\  {\bf 7}, 53 (2003)
[arXiv:hep-th/0302109];
H.~Ooguri and C.~Vafa,
``Gravity induced C-deformation,''
Adv.\ Theor.\ Math.\ Phys.\  {\bf 7}, 405 (2004)
[arXiv:hep-th/0303063].
}

\lref\AffleckXZ{
  I.~Affleck, M.~Dine and N.~Seiberg,
  ``Dynamical Supersymmetry Breaking In Four-Dimensions And Its
  Phenomenological Implications,''
  Nucl.\ Phys.\ B {\bf 256}, 557 (1985).
}

\lref\AffleckMK{
  I.~Affleck, M.~Dine and N.~Seiberg,
  ``Dynamical Supersymmetry Breaking In Supersymmetric QCD,''
  Nucl.\ Phys.\ B {\bf 241}, 493 (1984).
}

\lref\HatanakaRG{
  T.~Hatanaka, S.~V.~Ketov, Y.~Kobayashi and S.~Sasaki,
  ``Non-anti-commutative deformation of effective potentials in supersymmetric
  gauge theories,''
  arXiv:hep-th/0502026.
}

\lref\AmatiFT{
D.~Amati, K.~Konishi, Y.~Meurice, G.~C.~Rossi and G.~Veneziano,
``Nonperturbative Aspects In Supersymmetric Gauge Theories,''
Phys.\ Rept.\  {\bf 162}, 169 (1988).
}

\lref\KlemmYU{
J.~H.~Schwarz and P.~Van Nieuwenhuizen,
  ``Speculations Concerning A Fermionic Substructure Of Space-Time,''
  Lett.\ Nuovo Cim.\  {\bf 34}, 21 (1982);
P.~Bouwknegt, J.~G.~McCarthy and P.~van Nieuwenhuizen,
  ``Fusing the coordinates of quantum superspace,''
  Phys.\ Lett.\ B {\bf 394}, 82 (1997)
  [arXiv:hep-th/9611067];
S.~Ferrara and M.~A.~Lledo,
  ``Some aspects of deformations of supersymmetric field theories,''
  JHEP {\bf 0005}, 008 (2000)
  [arXiv:hep-th/0002084];
  D.~Klemm, S.~Penati and L.~Tamassia,
  ``Non(anti)commutative superspace,''
  Class.\ Quant.\ Grav.\  {\bf 20}, 2905 (2003)
  [arXiv:hep-th/0104190].
}

\lref\BrittoKG{
R.~Britto, B.~Feng and S.~J.~Rey,
  ``Deformed superspace, N = 1/2 supersymmetry and (non)renormalization
  theorems,''
  JHEP {\bf 0307}, 067 (2003)
  [arXiv:hep-th/0306215];
S.~Terashima and J.~T.~Yee,
  ``Comments on noncommutative superspace,''
  JHEP {\bf 0312}, 053 (2003)
  [arXiv:hep-th/0306237];
M.~T.~Grisaru, S.~Penati and A.~Romagnoni,
  ``Two-loop renormalization for nonanticommutative N = 1/2 supersymmetric  WZ
  model,''
  JHEP {\bf 0308}, 003 (2003)
  [arXiv:hep-th/0307099];
R.~Britto and B.~Feng,
  ``N = 1/2 Wess-Zumino model is renormalizable,''
  Phys.\ Rev.\ Lett.\  {\bf 91}, 201601 (2003)
  [arXiv:hep-th/0307165];
A.~Romagnoni,
  ``Renormalizability of N = 1/2 Wess-Zumino model in superspace,''
  JHEP {\bf 0310}, 016 (2003)
  [arXiv:hep-th/0307209];
O.~Lunin and S.~J.~Rey,
``Renormalizability of non(anti)commutative gauge theories with N = 1/2
supersymmetry,''
JHEP {\bf 0309}, 045 (2003)
[arXiv:hep-th/0307275];
D.~Berenstein and S.~J.~Rey,
``Wilsonian proof for renormalizability of N = 1/2 supersymmetric field
theories,''
Phys.\ Rev.\ D {\bf 68}, 121701 (2003)
[arXiv:hep-th/0308049];
M.~Alishahiha, A.~Ghodsi and N.~Sadooghi,
  ``One-loop perturbative corrections to non(anti)commutativity parameter  of N
  = 1/2 supersymmetric U(N) gauge theory,''
  Nucl.\ Phys.\ B {\bf 691}, 111 (2004)
  [arXiv:hep-th/0309037];
  I.~Jack, D.~R.~T.~Jones and L.~A.~Worthy,
  ``One-loop renormalisation of N = 1/2 supersymmetric gauge theory,''
  Phys.\ Lett.\ B {\bf 611}, 199 (2005)
  [arXiv:hep-th/0412009].
}

\lref\BrittoKGbis{
R.~Britto and B.~Feng,
``N = 1/2 Wess-Zumino model is renormalizable,''
Phys.\ Rev.\ Lett.\  {\bf 91}, 201601 (2003)
[arXiv:hep-th/0307165].
}

\lref\NovikovIC{
  V.~A.~Novikov, M.~A.~Shifman, A.~I.~Vainshtein and V.~I.~Zakharov,
  ``Supersymmetric Instanton Calculus: Gauge Theories With Matter,''
  Nucl.\ Phys.\ B {\bf 260}, 157 (1985)
  [Yad.\ Fiz.\  {\bf 42}, 1499 (1985)].
}

\lref\BilloJW{
  M.~Billo, M.~Frau, I.~Pesando and A.~Lerda,
  ``N = 1/2 gauge theory and its instanton moduli space from open strings in
  R-R background,''
  JHEP {\bf 0405}, 023 (2004)
  [arXiv:hep-th/0402160];
M.~Billo, M.~Frau, F.~Lonegro and A.~Lerda,
  ``N = 1/2 quiver gauge theories from open strings with R-R fluxes,''
  arXiv:hep-th/0502084.
}

\lref\AldrovandiHX{
  L.~G.~Aldrovandi, D.~H.~Correa, F.~A.~Schaposnik and G.~A.~Silva,
  ``BPS analysis of gauge field - Higgs models in non-anticommutative
  superspace,''
  Phys.\ Rev.\ D {\bf 71}, 025015 (2005)
  [arXiv:hep-th/0410256].
}



\Title {\vbox{\baselineskip12pt \hbox{\tt YITP-SB-05-12}
\hbox{\tt hep-th/0505077} 
}
} 
{\vbox{\centerline{
Instantons and Matter}\medskip\medskip \centerline{in ${\cal N}=1/2$ 
Supersymmetric Gauge Theory
}
}}
\centerline{ Simone Giombi, Riccardo Ricci,} \centerline{
Daniel Robles-Llana and Diego Trancanelli \foot{E-mails: {\tt
sgiombi, rricci, daniel, dtrancan@insti.physics.sunysb.edu}}}
\medskip
\centerline{\it C.N. Yang Institute for Theoretical Physics,}
\centerline{\it State University of New York at Stony Brook,}
\centerline{\it Stony Brook, NY 11794-3840, USA}

\bigskip
\bigskip
\bigskip
\bigskip

We extend the instanton calculus for ${\cal N}=1/2$ $U(2)$ supersymmetric gauge theory by including one massless flavor. We write the equations of motion at leading order in the coupling constant and we solve them exactly in the non(anti)commutativity parameter $C$. The profile of the matter superfield is deformed through linear and quadratic corrections in $C$. Higher order corrections are absent because of the fermionic nature of the back-reaction. The instanton effective action, in addition to the usual 't Hooft term, includes a contribution of order $C^2$ and is ${\cal N}=1/2$ invariant. We argue that the ${\cal N}=1$ result for the gluino condensate is not modified by the presence of the new term in the effective action.  
\noindent

\Date{\newdate}

\baselineskip16pt

\newsec{Introduction}

Non-anticommutative superspace \KlemmYU\  has been thoroughly 
investigated
 recently \SeibergYZ, the interest being in part motivated by a
surprising connection with string dynamics in non trivial
backgrounds. Another reason of interest is the study of theories with 
reduced supersymmetry. 

In non-anticommutative superspace half of the
superspace fermionic variables are promoted to elements of a
Clifford algebra. The complete set of anti-commutation relations
is then 
\eqn\deformed{\{\theta^{\a},\theta^{\b}\}=C^{\a\b}
\;\;\;~~~~~~\{\bar{\theta}^{\dot{\a}},\bar{\theta}^{\dot{\b}}\}=0
 \;\;\;~~~~~~\{\theta^{\a},\bar{\theta}^{\dot{\b}}\}=0,}
where $C^{\a\b}$ is the deformation parameter.
At the level of the
supersymmetry algebra the only modification due to the {\it
chiral} deformation $C^{\a\b}$ is in the
anti-commutator of the {\it anti-chiral} supercharges
\eqn\susy{\{\bar{Q}_{\dot{\a}},\bar{Q}_{\dot{\b}}\}=
-4C^{\a\b}\sigma^{\mu}_{\a\dot{\a}}\sigma^{\nu}_{\b\dot{\b}}{\partial^{2}\over{\partial
y^{\mu}\partial y^{\n}}}} where
$y^{\mu}=x^{\mu}+i\theta^{\a}\sigma^{\mu}_{\a\dot\a}\bar{\theta}^{\dot{\a}}$.
The amount of supersymmetry left is therefore $\cal{N}$=1/2 and
corresponds to the unbroken supersymmetry generators $Q_{\a} $.
 The non trivial anti-commutator of the $\theta^\a$ variables breaks
 also the Lorentz group $SU(2)_L\times SU(2)_R$ to $SU(2)_R$. One can write down 
 a super Yang-Mills theory lagrangian which is invariant under ${\cal N}=1/2$ supersymmetry only
\eqn\prima{{\cal L}={1\over g^2}{\rm
Tr} \left(-{1\over 8}F_{\m\n}F^{\m\n}-{i\over 2}\bar
\l \bar\s^{\m}{\cal D}_{\mu}\l+{1\over 4}D^2-iC^{\m\n}F_{\m\n}\bar\l 
\bar
\l+{|C|^2\over 4}(\bar\l \bar\l)^2\right).}
 It is not Hermitian and contains
 operators of dimensions 5 and 6 but still it is
 renormalizable \BrittoKG.

 As already remarked, a motivation for the study of theories in $C$-deformed superspace is 
given by
 their connection with strings propagating in geometries with non 
trivial background.
  More precisely,  the superspace deformation \deformed\  is
 related to $\cal N$=2 superstrings in Euclidean $R^4$ with
 self-dual graviphoton field-strength background
 $F^{\dot{\a}\dot{\b}}=0$ \OoguriQP\SeibergYZ\BerkovitsKJ. The 
parameter $C^{\a\b}$
 and the selfdual field-strength are related as
 $({\a'})^2F^{\a\b}=C^{\a\b}$ \foot{For string theory related aspects see also \BilloJW.}.

Perturbative analysis of Wess-Zumino models and ${\cal N}=1/2$
super Yang-Mills were performed in \SeibergYZ\BrittoKG\ while
instanton configurations were considered in
\ImaanpurJJ\ImaanpurIG.  The presence of the deformation
parameter $C^{\a\b}$ alters the usual instanton profile. A nice
feature of the $C$-deformed background is that it is fully solvable in the
deformation: The equations of motion can be solved iteratively in
the deformation parameter, the iteration in $C$ terminating after
a finite number of steps due to the fermionic nature of the
back-reaction. The simplest non trivial case to study is
 when the gauge group is $U(2)$ \ImaanpurJJ\GrassiQK\BrittoUV.
  The {\it anti-self-dual } instanton configuration 
coincides with the familiar 't Hooft-Polyakov
  instanton while the {\it self-dual} equations 
of motion are modified by the presence
  of a fermionic source
  \eqn\basic{F_{\m\n}^{+}+{i\over 2}C_{\m\n}{\bar\l}{\bar\l}=0.}
  The only novelty here is in the $U(1)$ component 
of the gauge field which is linear in the deformation and
quadratic in the Grassmann collective coordinates. The gluino
zero-modes and the topological charge do not receive corrections
in $C$. Generalization to $U(N)$ gauge groups appeared in
\BrittoUV.

 A natural extension for obtaining  more
realistic theories is to include flavors by coupling   ${\cal
N}=1/2$ $U(2)$ super Yang-Mills to a matter superfield. The ${\cal
N}=1/2$ lagrangian in the presence of chiral superfields has been
given in \ArakiSE.

Adding flavors  is useful because it allows one to work 
consistently in a
semi-classical approximation by choosing the Higgs vacuum expectation value suitably 
larger than the strong dynamics scale of the gauge theory. This is the only regime where instanton methods are reliable \foot{For a comprehensive review on these aspects see \DoreyIK.}.

On the other hand, the incorporation of
fundamental matter in instanton calculus substantially modifies the nature of the
problem. In particular, it is known that  
additional zero-modes for the fundamental fermions appear 
in the Higgs phase and
that the equations of motion can no longer be solved exactly but
only order by order in $g^2$. The instanton configuration is
therefore only an {\it approximate} solution.  At leading order
in the coupling constant the effects of matter are captured by
additional terms in the instanton effective action which depend
on the instanton size $\rho$, the asymptotic value  of the Higgs
field, and the extra fermionic collective coordinates
in the matter sector. The presence of an effective potential
which depends on the instanton moduli reflects the fact that we
are not fully solving the equations of motion.

In this paper we study instanton solutions of  ${\cal N}=1/2$ $U(2)$ super Yang-Mills 
with
one massless flavor \foot{Monopoles and vortices in a similar setting have been studied in \AldrovandiHX.}.  The matter sector is solved at leading order 
in perturbation theory by expanding around
the pure ${\cal N}$=1/2 super Yang-Mills background.  The new ingredients 
are the deformed $U(1)$ connection and $C$-dependent Yukawa-like interaction terms. The strategy 
is then  to reduce the equation of motions  to non homogeneous
Dirac and Laplace equations in the usual $SU(2)$ undeformed
instanton background. These equations are fully solved in $C$
 at the given leading order in $g$. 
We will see that the corrections to
the matter fields will be only linear and quadratic in the
deformation. Corrections of order $C^3$ do not appear due to the
Grassmannian nature of the iteration.
We finally
substitute the solutions in the action and evaluate their surface
contribution to obtain the instanton effective action.  The usual  't Hooft effective action  gets a new contribution which is quadratic in $C$ and quartic in the supersymmetry and superconformal collective coordinates.  Still the presence of the new term does not modify the gluino condensate.

This paper is organized as follows: In sections 2 and 3 we review
the ${\cal N}=1/2$ superalgebra, the construction of the lagrangian, and the  supersymmetry transformations of the
fields. In section 4 we 
present the 
equations of motion. In section 5 we give the systematics of
the expansion in the coupling constant while in section 6  we 
solve the equations of
motion at each order in the deformation parameter $C$. 
In sections 7 and 8 we
obtain the instanton effective action and we discuss its supersymmetry properties.
Finally, in section 9 we argue that the gluino condensate is not affected by the deformation.  Notation and details of the calculations are collected in a series of  
Appendices.



\newsec{Review of ${\cal N}=1/2$ supersymmetric theories}
Four dimensional ${\cal N}=1/2$ supersymmetric theories \SeibergYZ\ arise when half of the fermionic coordinates of superspace $\theta^\alpha$ obey a Clifford algebra
\eqn\thetas{
\{\theta^\alpha , \theta^\beta \}= C^{\a\b} ~.}
The non-trivial anticommutator \thetas\ implies that functions of $\theta$ should be Weyl ordered. This is achieved by introducing a fermionic star product
\eqn\stpr{
f(\theta) \star g(\theta) = f(\theta) \exp\left(-{C^{\a\b}\over 2}\overleftarrow{\partial\over\partial\theta^\alpha}
 \overrightarrow{\partial\over\partial\theta^\beta}\right) g(\theta)~.}
The rest of the coordinates obey their usual (anti)commutation relations
\eqn\resofac{
[ y^\m , y^\n ] = [ y^\m , \theta^\a ] = [y^m , \bar\theta^{\dot\a} ] =0 }
where, in order to preserve the above relations and be able to properly define chiral and antichiral superfields, one has to work with the chiral coordinate $y^\m = x^\m + i \theta^\a \s^\m_{\a\dot\a}\bar\theta^{\dot\a}$. This means that the supercharges take the form
\eqn\supchar{\eqalign{
Q_\a &= {\partial \over \partial\theta^\a} \cr
\bar Q_{\dot\a} &= -{\partial\over \partial\bar\theta^{\dot\a}} + 2i\theta^\a \partial_{\a\dot\a} }}
and that the supersymmetry algebra is deformed to
\eqn\susyalgebra{\eqalign{
\{Q_\a , Q_\a \} &=0 \cr
\{ \bar Q_{\dot\a} , Q_\a \} &=2 i \partial_{\a\dot\a} \cr
\{ \bar Q_{\dot\a} , \bar Q_{\dot\b} \}&= - 4 C^{\a\b}\partial_{\a\dot\a} \partial_{\b\dot\b}.}}
Therefore, only half of the initial supersymmetries are realized linearly. Furthermore the rest of the (anti)commutators of the ${\cal N}=1$ superconformal algebra which contain products of $\theta$'s are deformed, and the generators involved therein are no longer symmetries of the theory. The symmetry algebra that remains is the ${\cal N}=1/2$ super Poincare algebra 
\eqn\onehalfsuper{\eqalign{&
[\bar M_{\dot\a\dot\b} , \bar M_{\dot\gamma\dot\d}]=\e_{\dot\a(\dot\gamma}\bar M_{\dot\d)\dot\b}+\e_{\dot\b(\dot\gamma}\bar M_{\dot\d)\dot\a} \cr
 & [\bar M_{\dot\a\dot\b} , P_{\dot\gamma\dot\d}]=4P_{\g(\dot\a}\e_{\dot \b)\dot\g} \cr &
[\bar M_{\dot\a\dot\b} , \bar S_{\dot\gamma}] = \e_{(\dot\a\dot\g} \bar S_{\dot\b)} \cr &
[P_{\a\dot\a},\bar S_{\dot\b}] = 2i \e_{\dot\a\dot\b}Q_\a \cr &
[\tilde D,P_{\a\dot\a}] =-iP_{\a\dot\a} \cr &
[\tilde D,\bar S_{\dot\a}] =i\bar S_{\dot\a}~,}}
where the dilatation operator $\tilde D$ takes the form $\tilde D\equiv D-{1\over 2}R$ and accounts for the non trivial R-charge of $C$, which is given by $-2$.

One can write lagrangians in non(anti)commutative superspace. Note that the deformation \thetas\ is chiral and therefore these theories are well defined only in Euclidean space. The procedure is straightforward and consists in writing the usual ${\cal N}=1$ supersymmetric action but replacing all products by star products. We do this in the next section.


\newsec{The lagrangian of ${\cal N}=1/2$ super Yang-Mills with matter}

In this paper we will be interested in ${\cal N}=1/2$ $U(2)$ super Yang-Mills coupled to fundamental matter, a theory which has been considered in  \ArakiSE. The fundamental quark flavor corresponds to one chiral field in the ${\bf 2}$ representation of
$U(2)$ and one chiral field in the $\bar {\bf 2}$ representation of
$U(2)$. By writing the usual $F$- and $D$-terms for the usual undeformed ${\cal N}=1$ theory and replacing ordinary products with star products \stpr\ everywhere we obtain 
\eqn\lagrangian{\eqalign{ & {\cal L} = \int d^4\theta~~ \left[
\Phi^\dagger \star e^V_{\star} \star \Phi ~ + ~\tilde \Phi \star
e^{-V}_{\star} \star \tilde \Phi^\dagger \right]~+\cr & \hskip+2cm
+i \tau~ \int d^2 \theta ~{\rm Tr}~W^\alpha \star W_\alpha
-i\bar\tau\int d^2\bar\theta ~{\rm Tr}~W^\dagger_{\dot\a} \star
W^{\dagger\dot\a}\,,}} where the field-strengths are computed from
the usual definition, but replacing ordinary products with star
products as well. 

As pointed out in previous literature, there are some difficulties
concerning the definition of vector \SeibergYZ\ and antichiral
superfields \ArakiSE. These arise because of the fact that one would like
infinitesimal gauge transformations
\eqn\gaugetransformations{\eqalign{ & \delta \Phi = -i \Lambda
\star \Phi\,,~~~~~~~~~~~~~\delta\tilde \Phi =i
\tilde\Phi\star\Lambda\,,\cr & \delta \Phi^\dagger = i\Phi^\dagger
\star \Lambda^\dagger\,,
~~~~~~~~~~~\delta\tilde\Phi^\dagger=-i\Lambda^\dagger \star
\tilde\Phi^\dagger\,,\cr & \delta e^V= -i\Lambda^\dagger \star
e^V_{\star} ~ + ~ i e^{V}_{\star} \star\Lambda}} to act as
ordinary undeformed transformations on the component fields. This
problem is surmounted by modifying the definition of the vector
(in the Wess-Zumino gauge)  and antichiral matter
superfields from the undeformed case as
\eqn\vectorantichiral{\eqalign{ & V(y,\theta,\bar\theta)=-\theta
\sigma^\mu \bar\theta A_\mu (y) + i
\theta\theta\bar\theta\bar\lambda (y) -i\bar\theta\bar\theta
\theta^\alpha \left(\lambda_\alpha (y) +{1\over4} \epsilon_{\alpha
\beta}C^{\beta\gamma}\sigma_{\gamma\dot\gamma}^{\mu}
\{\bar\lambda^{\dot\gamma} ,A_\mu \}(y) \right) +\cr
&\hskip 3cm +
{1\over2}\theta\theta\bar\theta\bar\theta \left(D(y)-i\del_{\mu}A^{\mu}(y)\right)
\cr & \Phi^\dagger(\bar{y},\bar\theta)= A^\dagger (\bar y) +
\sqrt{2} \bar\theta\psi^\dagger(\bar y)
+\bar\theta\bar\theta\left(F^\dagger(\bar y) + i C^{\mu\nu}
\del_{\mu}( A^\dagger A_{\nu})(\bar y) -
{1\over4}C^{\mu\nu}A^\dagger A_{\mu} A_{\nu}(\bar y)\right)\, \cr
& \tilde\Phi^\dagger (\bar y,\bar\theta)=\tilde A^\dagger (\bar y)
+\sqrt{2} \bar\theta \tilde \psi^\dagger (\bar y)+
\bar\theta\bar\theta\left(\tilde F^\dagger(\bar y)
-iC^{\mu\nu}\del_\mu( A_\nu \tilde A^\dagger)(\bar y)-{1\over
4}C^{\mu\nu} A_{\mu}A_\nu\tilde A^\dagger(\bar y)\right)\, }}
where $y^\mu =x^\mu + i\theta \sigma^\mu \bar\theta$ is the
ordinary chiral coordinate and $\bar y^\mu = x^\mu - i\theta
\sigma^\mu\bar\theta$ is the antichiral coordinate. 

In order for ordinary gauge transformations 
to preserve the $C$-deformed Wess-Zumino gauge for $W_{\a}$, it can be shown that the usual chiral gauge parameter has to be modified as 
\eqn\gaugeparameters{\eqalign{ & \Lambda (y,\theta) = - \phi
(y)\cr & \Lambda^\dagger (\bar y, \bar\theta) = - \phi(\bar y)
-{i\over2} \bar\theta\bar\theta C^{\mu\nu} \{ \del_{\mu}\phi ,
A_{\nu} \} (\bar y).}}
One can then see that \gaugetransformations\ and \gaugeparameters\ reproduce ordinary gauge transformations for
all component fields, namely \eqn\ordinarygauge{\eqalign{ & \delta
A = i \phi A\,,~~~~~~~~~~~~~~\delta\psi = i\phi\psi\,,~~~~~~~~~~~~\delta F=i\phi F \,,\cr
& \delta\tilde A= -i\tilde A
\phi\,,~~~~~~~~~~~~\delta\tilde\psi=-i\tilde\psi\phi\,,~~~~~~~~~\delta
\tilde F=-i \tilde F \phi\,,\cr & \delta A_{\mu} = -2 \del_{\mu} \phi +
i[\phi,A_{\mu}]\,,~~~~~~\delta\lambda=
i[\phi,\lambda]\,,~~~~~~\delta\bar\lambda
=i[\phi,\bar\lambda]\,,~~~~~~\delta D=i[\phi,D]\,.}} 

Finally, 
by expanding the star products, one can write \lagrangian\ in components as 
\eqn\componentlagrangian{\eqalign{ & {\cal L} = {\rm Tr} \left(
-{1\over 2}F_{\mu\nu}F^{\mu\nu} -2i\bar\lambda
\bar\sigma^{\mu}{\cal D}_{\mu}\lambda + D^2 \right) +\cr &
\hskip+0.45cm + F^\dagger F -i\bar\psi \bar\sigma^{\mu}{\cal
D}_{\mu}\psi -{\cal D}_{\mu} A^\dagger {\cal D}^{\mu}A + g^2
A^\dagger D A +i{\sqrt 2 \over 2} g ( A^\dagger \lambda\psi -\bar\psi
\bar\lambda A) + 
\cr &\hskip+0.45cm 
+ \tilde F \tilde F^\dagger-i\tilde\psi 
\sigma^\mu {\cal D}_\mu \bar{\tilde\psi} -{\cal
D}_\mu \tilde A {\cal D}^\mu \tilde A^\dagger - g^2 \tilde A D
\tilde A^\dagger +i{\sqrt{2}\over 2}g\left(\tilde A\bar\l
\bar{\tilde\psi}-\tilde\psi\lambda\tilde A^\dagger\right)+\cr &
\hskip+0.45cm + 
{\rm Tr}\left( - i C^{\mu\nu} F_{\mu\nu}
\bar\lambda\bar\lambda +
{|C|^2\over4}(\bar\lambda\bar\lambda)^2\right) +\cr &
\hskip+0.45cm + 
i C^{\mu\nu}  A^\dagger F_{\mu\nu} F - {\sqrt 2\over 2} 
C^{\alpha\beta}\sigma^{\mu}_{\alpha\dot\alpha}{\cal D}_{\mu}
A^\dagger \bar\lambda^{\dot\alpha}\psi_{\beta} - {|C|^2\over 4}
  A^\dagger \bar\lambda\bar\lambda F-\cr &
\hskip+0.45cm -iC^{\mu\nu}\tilde F F_{\mu\nu}\tilde
A^\dagger-{\sqrt 2\over 2}  C^{\a\b}\sigma^{\mu}_{\a\dot\a}\tilde \psi_\b
\bar\lambda^{\dot\a}{\cal D}_\mu \tilde A^\dagger-{|C|^2\over 4}
\tilde F \bar\l\bar\l \tilde A^\dagger\,.}} 
In writing this expression we have
rescaled the vector multiplet as $V{\rightarrow} gV$ and also the deformation parameter as 
$C^{\m\n}{\rightarrow}{1\over g}C^{\m\n}$ so that the vector multiplet \vectorantichiral\
 is linear in $g$.
The first three lines
correspond to the undeformed F- and D-terms of the ${\cal N}=1$
lagrangian coupled to fundamental matter, whereas the last three
lines give the $C$-deformations of the F- and D-terms. In our
conventions $C^{\m\n}=C^{\a\b}\epsilon_{\b\g}
\s^{\m\n\,\g}_{\,\a}$ and $|C|^2=C_{\m\n}C^{\m\n}=4~ {\rm det} ~ C$. Moreover the covariant derivative is normalized as ${\cal D}_{\m}=\partial_{\m}+{i\over 2}A_{\m}$.

The lagrangian \componentlagrangian\ is invariant under a set of $C$-deformed
supersymmetries generated by acting with $Q_{\alpha}$ on the
various superfields of the theory. Because the deformed vector
superfield contains an additional $C$-dependent term in the
$\bar\theta^2 \theta$ component the supersymmetry transformation
of $\lambda$ is deformed. Similarly the supersymmetry
transformation of the $F^\dagger$ component of the antichiral
superfields also contains an extra $C$-dependent term. By carrying out the usual procedure one finds the supersymmetries of the theory, which are given by
\eqn\susy{\eqalign{ & \delta
A= \sqrt 2 \e \psi\,,~~~~~~\delta \psi_\alpha =\sqrt 2 \e_\alpha
F\,,~~~~~~~~\delta F=0\cr & \delta
A^\dagger=0\,,~~~~~~~~~~~\delta\bar\psi^{\dot\alpha}=-i\sqrt 2 {\cal
D}_{\mu} A^\dagger (\e\sigma^\mu)^{\dot\alpha}\cr & \delta
F^\dagger= -i\sqrt 2 {\cal D}_{\mu} \bar\psi\bar\sigma^{\mu}\e-i
A^\dagger\e\lambda+C^{\mu\nu}\left[\del_{\mu}(A^\dagger
\e\sigma_{\nu}\bar\lambda)-{i\over2}(
A^\dagger\e\sigma_{\nu}\bar\lambda)A_{\mu}\right]\cr & \delta
A_{\mu}=-i\bar\lambda\bar\sigma_{\mu}\e\,,~~~~~~\delta
D=-\e\sigma^{\mu}{\cal D}_{\mu}\bar\lambda \cr & \delta
\lambda_{\alpha}=i\e_\alpha
D+(\sigma^{\mu\nu}\e)_{\alpha}\left(F_{\mu\nu}+{i\over2}
C_{\mu\nu}\bar\lambda\bar\lambda\right)\,,~~~~~\delta\bar\lambda
^{\dot\alpha}=0\,,}} with similar expressions for the tilded
fields. Note that we only wrote the chiral supersymmetries, as the anti-chiral ones are explicitely broken by the deformation. In addition, we see that only the tranformation of the antichiral auxiliary field and of the chiral gluino are modified from the usual case.


\newsec{The equations of motion}
The preliminary step before solving the equations of motion
consists in eliminating the auxiliary fields from the action. 
This simplifies the lagrangian considerably. The $F^\dagger$
equation of motion implies $F=0$, which eliminates two of the
$C$-deformed D-terms. Using the equations of motion for $D$, which
are the same as in the undeformed case, we can write the
lagrangian as 
\eqn\onshelllagrangian{\eqalign{ & {\cal L} = {\rm
Tr} \left( -{1\over 2}F_{\mu\nu}F^{\mu\nu} -2i\bar\lambda
\bar\sigma^{\mu}{\cal D}_{\mu}\lambda \right) -\cr & \hskip+0.45cm
-{\cal D}_{\mu} A^\dagger {\cal D}^{\mu}A
-i\bar\psi\bar\sigma^{\mu}{\cal D}_{\mu}\psi +i{\sqrt 2\over 2} g\left(A^\dagger
\lambda\psi -\bar\psi\bar\lambda A\right)  + {1\over 4} g^2 \left(A^\dagger
T^a A-\tilde A T^a \tilde A^\dagger\right)^2 -\cr & \hskip+0.45cm -{\cal
D}_\mu \tilde A {\cal D}^\mu \tilde A^\dagger
-i\tilde \psi\sigma^\mu{\cal D}_\mu \bar{\tilde\psi}+i{\sqrt 2\over 2}
g\left(\tilde A\bar\lambda\bar{\tilde\psi}-\tilde\psi\l\tilde
A^\dagger\right)+\cr & \hskip+0.45cm + {\rm Tr}\left( - i
 C^{\mu\nu} F_{\mu\nu} \bar\lambda\bar\lambda +
 {|C|^2\over4}(\bar\lambda\bar\lambda)^2\right) -\cr &
\hskip+0.45cm -{\sqrt 2\over 2} 
C^{\alpha\beta}\sigma^{\mu}_{\alpha\dot\alpha}{\cal D}_{\mu}
A^\dagger \bar\lambda^{\dot\alpha}\psi_{\beta} -{\sqrt 2\over 2} 
C^{\a\b}\sigma^\m_{\a\dot\a} \tilde\psi_\b \bar\l^{\dot\a}{\cal
D}_\m \tilde A^\dagger\,.}} 

We are now ready to write the
equations of motion for the different fields. They read \eqn\eom
{\eqalign{ & A^\n:\;\;\; {\cal D}_\mu \left( F^{\mu\nu} + i
C^{\mu\nu} \bar\lambda\bar\lambda \right) + ig \{
\bar\lambda_{\dot\alpha}\bar\sigma^{\nu\;\dot\alpha\beta} ,
\lambda_{\beta} \} + g
\bar\sigma^{\nu\;\dot\alpha\beta}\psi_{\beta}\bar\psi_{\dot\alpha}
-2ig A^\dagger {\cal D}^{\nu}A + \cr & \hskip 2cm + {\sqrt 2\over 2} i 
C^{\alpha\beta} \sigma^{\nu}_{\alpha\dot\alpha}
(\bar\lambda^{\dot\a} \psi_{\beta} A^\dagger+\tilde\psi_{\b}
\bar\lambda^{\dot\a}\tilde A^\dagger) - g
\bar\sigma^{\nu\;\dot\alpha\beta}\bar{\tilde
\psi}_{\dot\a}\tilde\psi_{\b}
 -2ig \tilde A {\cal D}^{\nu}\tilde A^\dagger = 0\,,
\cr & \l:\;\;\; 
\bar\sigma^{\mu\;\dot\alpha\alpha} {\cal D}_{\mu}
\lambda_{\alpha} + \bar\l^{\dot\a} \left( C_{\mu\nu}F^{\mu\nu} +i 
{|C|^2\over 2} \bar\lambda\bar\lambda\right)+ {1\over \sqrt 2} g A
\bar\psi^{\dot\a} -{1\over \sqrt 2} g
 \bar{\tilde\psi}^{\dot\a} \tilde A-
\cr & \hskip 2cm
 -i {1\over \sqrt 2}  C^{\alpha\beta}
\sigma^{\mu\;\dot\a}_{\alpha} [\psi_{\beta}{\cal D}_{\mu}
A^\dagger+({\cal D}_{\mu} \tilde A^\dagger)\tilde \psi_{\b}]=0\,,
\cr & \bar\l:\;\;\;
\sigma^{\mu}_{\alpha\dot\alpha}{\cal D}_{\mu}
\bar\lambda^{\dot\alpha} + {g\over \sqrt 2}  \psi_{\a} A^\dagger -
{g\over \sqrt 2} \tilde A^\dagger \tilde\psi_{\a}=0\,,
\cr & A: \;\;\;
{\cal D}^2 A + i{\sqrt 2\over 2} g \lambda \psi + {1\over 2}
g^2 (A^\dagger T^a A-\tilde A T^a \tilde A^\dagger) T^a A +\sqrt 2
 C^{\alpha\beta} \sigma^{\mu}_{\alpha\dot\alpha}{\cal
D}_{\mu}(\bar\lambda^{\dot\alpha}\psi_{\beta})=0 \,, \cr &
A^\dagger: \;\;\;{\cal D}^2 A^\dagger -i{\sqrt 2\over 2} g \bar\psi
\bar\lambda+ {1 \over 2} g^2 (A^\dagger T^a A- \tilde A T^a \tilde
A^\dagger)A^\dagger T^a =0\,, \cr & \psi:
\;\;\;\bar\sigma^{\mu\;\dot\alpha\alpha} {\cal D}_{\mu}
\psi_{\alpha} + {\sqrt 2\over 2} g \bar\lambda^{\dot\alpha}A=0\,, \cr &
\bar\psi: \;\;\;\sigma^{\mu}_{\alpha\dot\alpha} {\cal D}_{\mu}
\bar\psi^{\dot\alpha} - {\sqrt 2\over 2} g A^\dagger \lambda_{\alpha} -i
{\sqrt 2\over 2}  C^{\beta}_{\alpha}\sigma^{\mu}_{\beta\dot\beta}({\cal
D}_{\mu} A^\dagger) \bar\lambda^{\dot\beta}=0\,.  }} and similar
equations for $\tilde A$, $\tilde A^\dagger$, $\tilde \psi$, and
$\bar{\tilde \psi}$.



\newsec{Expansion in $g$ of the equations of motion}
In usual ${\cal N}=1$ instanton calculations with matter
fields there are no exact solutions to the equations of motion
when the lowest components of the fundamental chiral fields
acquire vacuum expectation values. The next best thing to do, if
one does not want to use constrained instantons \AffleckMP, is to solve the
equations approximately to leading order in the coupling constant.
To that purpose one solves for the matter fields around the
approximate solutions \foot{A superscript 
index on the left indicates the order of the expansion in the coupling constant. Later 
we will use superscripts on the right to indicate the order of the expansion in the $C$-deformation.}\eqn\approximate{
F_{\m\n}^+=0~~~~~~~{^{(0)}}{\cal D}^2 A=0} where the
covariant Laplace equation is taken in the instanton background.
By appropriate partial integration of the kinetic term for the
chiral scalars one can capture the leading effect due to the
non-vanishing Yukawa terms.

We will demonstrate that the same procedure holds if
one expands around the deformed instanton solutions of ${\cal
N}=1/2$ super Yang-Mills. The leading effect of this deformation,
for the $U(2)$ case, is to modify the zero-mode equation for the
fundamental fermions and the covariant Laplace equations for the
scalars.


\subsec{Systematics of the expansion in the coupling constant}
As a first step we proceed to set up a systematic expansion for solving 
the equations of motion order by order in the coupling constant. 
To lowest order the equations in the pure gauge sector are 
\eqn\lowest{\eqalign{
& A^\n:\;\;\; {\cal D}_\mu \left( F^{\mu\nu} + i 
C^{\mu\nu} \bar\lambda\bar\lambda \right)  = 0
\cr & \l:\;\;\; \bar\sigma^{\mu\;\dot\alpha\alpha} {\cal D}_{\mu}
\lambda_{\alpha} + \bar\l^{\dot\a} \left( C_{\mu\nu}F^{\mu\nu} +i 
{|C|^2\over 2} \bar\lambda\bar\lambda\right)=0
\cr & \bar\l:\;\;\;\sigma^{\mu}_{\alpha\dot\alpha}{\cal D}_{\mu}
\bar\lambda^{\dot\alpha}=0}}
whose solution was studied in \ImaanpurJJ\GrassiQK\BrittoUV.  
For the $U(2)$ theory that we are considering in this paper the solutions
 can be written in the form
\eqn\sollowga{\eqalign{
&A_{\a\dot\a}=g^{-1}\left(A_{\a\dot\a}^{SU(2)}+A_{\a\dot\a}^{C}\right)\cr
& \bar\l^{\dot\a}=g^{-1/2}\bar\l^{SU(2)\,\dot\a}\,,~~~~~\l_\a =0\,}}
where $A^{SU(2)}$ and $\bar\l^{SU(2)}$ are the usual solutions for the undeformed $SU(2)$ theory
 \eqn\instanton{\eqalign{& A_{\a\dot\a}^{SU(2)\, \dot a\dot b}(x,x_0
,\rho) ={-2i\over (x-x_0)^2+\rho^2}\left(\d_{\dot \a}^{\dot
a}(x-x_0)_{\a}^{\dot b}+\d_{\dot\a}^{\dot b}(x-x_0)_{\a}^{\dot
a}\right) \cr
&  \bar\lambda_{\dot\a}^{SU(2)\,\dot a\dot b}(x,x_0,\rho,\bar\xi)
 = {8i\r^2\over [(x-x_0)^2+\r^2]^2}\left(\d_{\dot\a}^{\dot a }\d_{\dot\b}^{\dot b }+\d_{\dot\a}^{\dot b}\d_{\dot\b}^{\dot a }\right)\bar\xi^{\dot\b} }} where $\bar\xi^{\dot \b}=\bar\zeta^{\dot\b}+x_{\;\a}^{\dot\b}\eta^{\a}$  is the supersymmetric fermionic collective coordinate. $A^C$ is a $C$-dependent correction which affects only the $U(1)$ subgroup of $U(2)$. It is given by \GrassiQK
\BrittoUV
\eqn\co{
A_{\a\dot\a}^{C}(x,x_0,\r,\bar\zeta,\eta)=-4iC_\a^{\;\b}\partial_{\b\dot\a}(2\bar\zeta^2K_1
+ 2\rho^2\eta^2 K_2 + 4\rho^2 K_3)~,}
with
\eqn\K{\eqalign{ & K_{1}(x,x_{0},\r) = {(x-x_{0})^{2}\over
[(x-x_{0})^{2}+\r^{2}]^{2}} - {2\over (x-x_{0})^{2}+\r^{2}} \cr
& K_{2}(x,x_{0},\r) = {(x-x_{0})^{2}\over [(x-x_{0})^2 +
{\r^2}]^2} + {1\over (x-x_{0})^2 + {\r^2}} \cr &
K_3(x,x_{0},\r) = {{\bar\zeta_{\dot\a}
\eta_\a}(x-x_{0})^{\dot\a\a}\over [(x-x_{0})^2 + \r^2]^2}\,. }}


Explicitly this deformed connection reads 
\eqn\erterd{ A_{\a\dot\a}^C=A_{\a\dot\a}^{(\bar\zeta^2)}+A_{\a\dot\a}^{(\eta^2)}+A_{\a\dot\a}^{(\bar\zeta\eta)}} where \foot{For simplicity, we set from this moment $x_0=0$.}
\eqn\explicitconnection{\eqalign{ 
& A_{\a\dot\a}^{(\bar\zeta^2)}=16iC_\a^{~\b} x_{\b\dot\a}{x^2+3\r^2 \over (x^2+\r^2)^{3}}\bar\zeta^2 \cr
& A_{\a\dot\a}^{(\eta^2)}=-32i\r^2\,C_\a^{~\b} x_{\b\dot\a}{x^2\over (x^2+\r^2)^3}\eta^2 \cr
& A_{\a\dot\a}^{(\bar\zeta\eta)}=-32i\r^2C_{\a}^{~\b}\bar\zeta_{\dot\g}\eta_{\g}{2x^{\dot\g\g}x_{\b\dot\a}-\delta_\b^\g \delta^{\dot\g}_{\dot\a}(x^2+\r^2)\over (x^2+\r^2)^3}.}}

From inspection of the action and of the equations of motion 
one arrives at the following standard $g$-expansion of the various fields \DoreyIK
\eqn\expansion{\eqalign{
& ~~~~~~~~~~~~~~~~~~~~~~~~~A_{\m}=g^{-1}~{^{(0)}}A_{\m}+g~{^{(1)}}A_{\m}+\ldots \cr 
& \bar\l=g^{-1/2}~{^{(0)}}\bar\l+g^{3/2}~{^{(1)}}\bar\l+\ldots 
~~~~~~ \l=g^{1/2}~{^{(0)}}\l+g^{5/2}~{^{(1)}}\l+\ldots \cr
&\bar\psi=g^{-1/2}~{^{(0)}}\bar\psi+g^{3/2}~{^{(1)}}\bar\psi+\ldots 
~~~~~\psi=g^{1/2}~{^{(0)}}\psi+g^{5/2}~{^{(1)}}\psi+\ldots \cr
&A^\dagger = g^0 ~{^{(0)}}A^{\dagger}+g^2 ~{^{(1)}}A^{\dagger}+\ldots 
~~~~~~~~A=g^0 ~{^{(0)}}A+g^2 ~{^{(1)}}A+\ldots
}}
We remark here that, as in the undeformed case, there are no non-trivial 
solutions for the equation for the $\psi$ zero-modes
 \foot{Writing $\bar{{\cal D}} =\bar{{\cal D}}^{(0)}+A^C$ and $\psi=\psi^{(0)}+\psi^{(C)}$ the equation at order $C^0$ gives just $\bar{{\cal D}}^{(0)}\psi^{(0)}=0$, which implies $\psi^{(0)}=0$ because $\bar{{\cal D}}^{(0)}$ is just the usual Dirac operator in the instanton background, which has no zero-modes. Then at order $C$ one has $\bar{{\cal D}}^{(0)}\psi^C=0$ which in turn gives $\psi^C=0$.}
\eqn\ghggjhn{\bar\sigma^{\mu\;\dot\alpha\alpha} {\cal D}_{\mu}
\psi_{\alpha}
=0 .}
Therefore the $g$ expansion for $\psi$ starts at order $1/2$. Usually one drops this field because its contribution to the action through the term $ g A^\dagger \lambda\psi $ is not leading order. However we need to include $\psi$ in the present case because it contributes to the leading order effective action through the $C$-dependent Yukawa-like interaction. 
 


\subsec{${\cal N}=1/2$ with one flavor}

In the absence of an instanton background, the ${\cal N}=1/2$ theory 
has flat directions in field space. These directions 
are found by solving the
D-flatness equations \eqn\dflat{ D^a = \left( A^\dagger T^a A -
\tilde A T^a \tilde A^\dagger \right) = 0.} We denote the vacuum expectation values of $A$ and $\tilde A$ with $q$ and $\tilde q$.  They are solutions to
\dflat\ and read \eqn\dflasol{ q^{\a} = \left(\matrix{q \cr
0}\right)~~~~~~~~~~\tilde{q}_{\a} = \left(\matrix{\tilde q &
0}\right)} with $|q|^2 = |\tilde q |^2$.

Now, as soon as the instanton background is switched on the Dirac
operator for the fundamental fermions has zero-modes, which are
deformed through $C$-dependent terms. These feed on the right-hand
side of the covariant Laplace equation for $A^\dagger$. One then
ends up having to solve the coupled equations
\eqn\coupledw{\eqalign{ & {\cal D}^2 A=-{\sqrt 2\over 2} C^{\a\b}\s_{\a\dot\a}^{\m}{\cal D}_{\m}
(\bar\l^{\dot\a}\psi_{\b})\,,
\cr
& {\cal D}^2 A^\dagger =i{\sqrt 2\over 2} g \bar\psi \bar\l\,,
\cr & 
\bar{\cal D}^{\dot\a \a}
\psi_{\alpha} =- {\sqrt 2\over 2} g \bar\lambda^{\dot\alpha}A\,,\cr &
{\cal D}_{\a\dot\a} \bar\psi^{\dot\a}=i
{\sqrt 2\over 2}  C^{\;\beta}_{\alpha}\sigma^{\mu}_{\beta\dot\beta}({\cal
D}_{\mu} A^\dagger) \bar\lambda^{\dot\beta}\,,
}} 
where the covariant
derivative is with respect to the deformed connection given in
\sollowga.
The equations for the scalar fields must be supplemented with
the boundary conditions at infinity $A \rightarrow  q$, $\tilde A
\rightarrow \tilde q$. 
There are similar equations also for the tilded fields. 

We stress now that in our solutions we
set the matter auxiliary fields $F=F^\dagger=0$, and these conditions are preserved under a chiral supersymmetry transformation. For $F$ this is obvious, as the
transformation is just $\delta_\e F=0$. For $F^\dagger$ the order $g^0$ part of
the ${\cal N}=1/2$ transformation is proportional to the deformed equation of
motion for $\bar\psi$, and thus is also zero for our solution.
Similarly, the $D=0$ constraint is preserved on-shell.


\newsec{Solution to the leading order coupled equations} We saw in
Section 5 that in order to consistently couple the pure ${\cal
N}=1/2$ super Yang-Mills instanton to matter to leading order in the coupling constant 
one had to solve \coupledw. We rewrite them in two-component spinor notation, abandoning sigma matrices \foot{We refer to the Appendix A for our conventions.}
\eqn\coupled{\eqalign{ & {\cal D}^2 A=-h{\sqrt 2\over 2}  C^{\a\b}{\cal D}_{\a\dot\a}
(\bar\l^{\dot\a}\psi_{\b})
\cr
& {\cal D}^2 A^\dagger =i{\sqrt 2\over 2} g \bar\psi_{\dot\a} \bar\l^{\dot\a}
\cr & 
\bar{\cal D}^{\dot\a \a}
\psi_{\alpha} =- {\sqrt 2\over 2} g \bar\lambda^{\dot\alpha}A\cr &
{\cal D}_{\a\dot\a} \bar\psi^{\dot\a}=i h  
{\sqrt 2\over 2}  C^{\beta}_{\alpha}({\cal
D}_{\b\dot\b} A^\dagger) \bar\lambda^{\dot\beta},
}} 
where ${\cal D}$ is taken with respect to the
deformed instanton connection and reads explicitly
\eqn\DD{{\cal D}_{\a\dot\a}=\partial_{\a\dot\a}+{i\over 2}A^{SU(2)}_{\a\dot\a}
+{i k\over 2}A^{C}_{\a\dot\a}\equiv {\cal D}^{(0)}_{\a\dot\a}+{i k\over 2}A^{C}_{\a\dot\a}.
}

In \coupled\ and \DD\ we have introduced two book-keeping parameters $h$ and $k$. This helps in keeping track of the contributions coming from the $U(1)$ part of the connection and the new Yukawa-like $C$-dependent interaction. In the final results reported in the main text we set $k=1$ and $h=1$, as required by supersymmetry of the action  \onshelllagrangian. On the other hand, we keep $k$ and $h$ explicit in the Appendix B to stress the remarkable simplifications occurring when $k=1=h$. 

We will solve \coupled\ order by order
in $C$. To do so we write the different fields as an expansion in
$C$ \eqn\cexpan{\eqalign{ &
A=A^{(0)}+A^{(1)}+\ldots\,,~~~~~~~~~~\tilde A=\tilde
A^{(0)}+\tilde A^{(1)}+\ldots\cr &
A^\dagger=A^{\dagger(0)}+A^{\dagger(1)}+\ldots\,,~~~~~~~\tilde
A^\dagger=\tilde A^{\dagger(0)}+\tilde A^{\dagger(1)}+\ldots\,,\cr
& \bar\psi=\bar\psi^{(0)}+\bar\psi^{(1)}+\ldots\,,~~~~~~~~~~~
\bar{\tilde\psi}=\bar{\tilde\psi}^{(0)}+\bar{\tilde\psi}^{(1)}+\ldots}}
where the superscript counts the power of $C$.

The general strategy to solve \coupled\ is to recast them in the form of Poisson equations 
in the background of the undeformed $SU(2)$ instanton. Generically we will encounter equations of the form
\eqn\rttr{({\cal D}^{(0)})^2 ~ \Phi={\cal J}}
where ${\cal J}$ may contain contributions from the deformed connection $A^C$. The solution to \rttr\ is given by the sum of the solution to the homogeneous equation and a particular solution. The latter is obtained by inverting the laplacian with the appropriate propagator \AmatiFT , given in the Appendix A. 


\subsec{Zero-th order in $C$} At zero-th order in $C$ the equations to
solve are \eqn\zerothorder{\eqalign{ & ({\cal D}^{(0)})^{
2}A^{(0)}=0 \cr & ({\cal D}^{(0)})^{2}A^{\dagger(0)}=i{\sqrt 2\over 2} g
\bar\psi^{(0)}_{\dot\a}\bar\l^{\dot\a}
\cr & 
\bar{\cal D}^{(0)\dot\a \a}
\psi_{\alpha (0)} =- {\sqrt 2\over 2} g \bar\lambda^{\dot\alpha}A^{(0)} \cr & {\cal
D}_{\a\dot\a}^{(0)}\bar\psi^{\dot\a (0)}=0\,,}} and similarly for the tilded
fields. These equations are the usual ones that one encounters in ${\cal N}=1$ supersymmetric QCD and have 
already been studied in the literature (see for example \ShifmanMV \DoreyIK).

The details of our computations are contained in the Appendix B. Here we only present the final results.

The solution to the first equation in \zerothorder\ has to be subject to the boundary
condition $A\rightarrow q$ at infinity in order to satisfy \dflat. One finds that  
\eqn\covlapsol{A^{\dot a (0)}(x,\r)=q^{\a}{x^{\dot a}_{\a}\over \sqrt{x^2+\rho^2}}.
}
We remind that $q^{\alpha}$ is a constant two-components
vector which is set to $q^{\alpha}=q \delta^{\alpha}_1$ by the D-term constraint \dflat.

The next equation one can solve is the equation for the fundamental fermionic zero-mode 
$\bar\psi^{(0)}$. Its solution is 
\eqn\firstordermatter{
 \bar\psi^{(0)}_{\dot\a \dot a} (x,\r,{\cal K})=g^{-1/2}{\rho^2 \epsilon_{\dot a \dot\a}\over (x^2+\r^2)^{3/2}}{\cal K}\, ,}
where ${\cal K}$ is a fermionic fundamental collective coordinate.

Having obtained \firstordermatter\ one can solve the second 
equation in \zerothorder\ by inverting the covariant 
Laplacian and adding the solution of the homogeneous 
equation, which is essentially \covlapsol. This procedure yields 
\eqn\adagge{A^{\dagger (0)}_{\dot a}(x,\r,{\cal K},\bar\xi)=q^{\dagger}_{\a}{x^{\a}_{\dot a}\over 
\sqrt{x^2+\r^2}}-{\sqrt 2 \rho^2 {\cal K}\bar\xi_{\dot a}\over (x^2+\r^2)^{3/2}}\, ,}
where $\bar\xi^{\dot a}=
\bar\zeta^{\dot a}+x^{\dot a}_\a \eta^\a$ are the adjoint fermionic collective coordinates inherited from $\bar\l$.

Finally, the solution for the fundamental fermionic field $\psi^{(0)}$ is 
\foot{Note that in the usual undeformed case $\psi^{(0)}$ is taken to be zero, since  
$\bar{{\cal D}}^{(0)}$  has no zero-modes.}
\eqn\psiiii{\psi^{\a \dot a (0)}(x,\rho,\bar\xi)=
-g^{1/2}{2 \sqrt{2} i  \rho^2 \over (x^2+\rho^2)^{3/2}}
q^\a \bar\xi^{\dot a}\, .}
This equation is solved by acting on both sides of it with a covariant derivative 
${\cal D}_{\b\dot\a}^{(0)}$ and then solving the 
resulting Poisson equation \foot{We use ${\cal D}
_{\b\dot\a}^{(0)}\bar{\cal D}^{(0)\dot\a \a}=
-\d_{\b}^{\a}({\cal D}^{(0)})^2$. Note that 
there is no field-strenght piece because self-duality 
of the instanton background implies $F^{(0)\a}_{\;\;\;\;~\b}=0$.}. 
In this case there is no solution to the homogeneous 
equation because of the self-duality of 
the instanton background we have chosen.


Before moving to the first order, we briefly comment on how the structure of the solutions  is dictated by supersymmetry. This offers an alternative way to derive
these solutions. In a purely bosonic background, the bosonic zero modes  are \eqn\bos{
A^{(0) \dot a}={q^\a
x^{\dot{a}}_{\a}\over \sqrt{x^2+\rho^2}}~~~~~~~~~~~~~~
A^{\dagger (0)}_{\dot
a}= {{q_\a^{\dagger} x_{\dot{a}}^{\a}\over
\sqrt{x^2+\rho^2} }}.}  The solution for $\psi^{(0)}$ is  obtained
by applying the broken supercharge
$\bar{Q}_{\dot\alpha}$ with the collective coordinate
$\bar{\zeta}^{\dot \alpha}$ as supersymmetry parameter
\eqn\pillo{{\psi}^{\alpha\dot a}=\delta_{\bar Q}{
\psi}^{\alpha\dot a}\sim
\bar\zeta_{\dot\alpha}\left(\bar{{\cal D}}^{\dot{\alpha}\a} A\right)^{\dot{a}}\sim
{q^{\alpha}\bar\zeta^{\dot a}\over (x^2+\rho^2)^{3/2}}.}  
We saw in the previous paragraph that the origin of this solution is the Yukawa interaction in the $\psi$ equation since $\bar{\cal D}^{(0)\dot\a \alpha}$ has no normalizable zero-modes.  In the $SU(2)$ self-dual background there are two other
fundamental fermionic zero-modes which are generated by applying
the broken superconformal generator $S^{\alpha}$ with $\eta$ as
superconformal parameter. This amounts to replacing $\bar\zeta\rightarrow \bar\xi$ and thus gives \psiiii. The
antichiral fermionic solution can be found observing that, in
presence of a non-vanishing vacuum expectation value $q$ also the {\it chiral}
supersymmetry generators act not trivially. The corresponding
transformation parameter gets identified with the fundamental
fermionic collective coordinate ${\cal K}_{\alpha}$ and the fermionic
solution reads \eqn\pillo{{\bar
\psi}_{\dot{\alpha}\dot{a}}\equiv\delta_Q{\bar
\psi}_{\dot{\alpha}\dot{a}}\sim
{\cal K}^{\alpha}\left( {\cal D}_{\alpha\dot{\alpha}}A^{\dagger}\right)_{\dot a}\sim
{\epsilon_{\dot a \dot{\alpha}} \r^2 q^{\dagger}_{\alpha}{\cal K}^{\alpha}\over
(x^2+\rho^2)^{3/2}}.} This is the zero mode of the ${\cal D}^{(0)}_{\alpha \dot {\alpha}}$ operator given in \firstordermatter, once we identify ${\cal K}\equiv q^{\dagger}_{\a}{\cal K}^{\a}$. With the antichiral fermion switched on we generate a new contribution to
 $A^{\dagger}$   using the broken supercharge $\bar{Q}_{\dot \alpha}$
\eqn\third{A_{\dot{a}}^{\dagger new}=A_{\dot{a}}^{\dagger}+\delta_{\bar{Q}} A_{\dot{a}}^{\dagger}
=q^{\dagger}_{\a}{x^{\a}_{\dot{a}}\over
\sqrt{x^2+\rho^2}}-\bar{\zeta}_{\dot{a}}{\sqrt 2 \r^2 {\cal K}\over
(x^2+\rho^2)^{3/2}}.} As before $\bar\zeta\rightarrow \bar\xi$ applying
also the broken superconformal generator $S$.


\subsec{First order in $C$} We now move on to the first order in the deformation parameter.
The corresponding equations are
\eqn\firstorder{\eqalign{ & \bar{\cal D}^{\dot\a \a(0)}{\cal D}_{\a \dot\a}^{(0)} A^{(1)} +
igkA^{\dot\a\a (1)} {\cal D}^{(0)}_{\a\dot\a }A^{(0)} =
h \sqrt 2 C^{\a\b}{\cal D}^{(0)}_{\a\dot\a}
(\bar\l^{\dot\a}\psi_{\b}^{(0)})\cr & \bar{\cal D}^{\dot\a \a(0)}
{\cal D}_{\a \dot\a}^{(0)} A^{\dagger (1)} -igkA^{\dot\a\a (1)} 
{\cal D}^{(0)}_{\a \dot\a} A^{\dagger
(0)}=-i\sqrt 2 g \bar\psi_{\dot\a}^{(1)}\bar\l^{\dot \a}\cr &
\bar{{\cal D}}^{\dot\a \a (0)}\psi_{\a}^{(1)}+{ig\over 2} k A^{\dot\a \a (1)}\psi_{\a}^{(0)}=
-{\sqrt 2\over 2} g \bar\l^{\dot\a}A^{(1)}\cr &
{\cal D}_{\a\dot\a}^{(0)}\bar\psi^{\dot\a (1)}
-{ig\over 2} k A^{(1)}_{\a\dot\a} \bar\psi^{\dot\a (0)} =
 i h
{\sqrt 2\over 2}  C^{\;\beta}_{\alpha}({\cal
D}^{(0)}_{\b\dot\b} A^{\dagger(0)}) \bar\lambda^{\dot\beta}.}}
Again, the reader is referred to the Appendix B for details of the computations.

The first order correction to the fundamental scalar $A^{(1)}$ is 
quadratic in the Grassmannian collective coordinates and
turns out to be
\eqn\fundscala{\eqalign{
A^{(1) \dot a}
= & {2 \r^2 q_{\a} C^{\a\b}\over (x^2+\r^2)^{5/2}}
\Big[x^{\dot a}_{\b}\left(\bar\zeta^2-2\bar\zeta_{\dot \g}
x^{\dot\g \g}\eta_{\g}+
\r^2\eta^2\right)+
 2(x^2+\r^2)\bar\zeta^{\dot a}\eta_{\b}\Big].
}}

The equation for the spinor $\bar\psi^{(1)}$ can be converted 
into a Poisson equation by introducing a fermionic prepotential via the Ansatz 
\eqn\psiansatza{\bar\psi^{\dot \a (1)}=\bar{\cal D}^{\dot \a \b}\Psi^{(1)}_{\b}}
and solving for $\Psi^{(1)}$ \foot{In this case it is not convenient to act on the equation with a covariant derivative because this would produce a curvature piece since 
$\bar{\cal D}^{(0)\dot\b \a}{\cal D}
_{\a\dot\a}^{(0)}\sim
-\d_{\dot\b}^{\dot\a}({\cal D}^{(0)})^2+F^{(0)\dot \a}_{\;\;\;\;\dot \b}$.}. 
The final solution for $\bar\psi^{(1)}$ consists of two contributions proportional to ${\cal K}$
and $q$
\eqn\psiuno{\bar\psi^{(1)}=\bar\psi^{(1)}_{{\cal K}}
+\bar\psi^{(1)}_{q}}
where
\eqn\psiunosolution{
\eqalign{ &
(\bar\psi^{(1)}_{{\cal K}})^{\dot \a}_{\dot a}=g^{-1/2}
{4 {\cal K} \r^2 C_{\a}^{\; \b} x^{\dot \a \a}
\over (x^2+\rho^2)^{7/2}}\Big[x_{\b\dot a}\left(\bar\zeta^2-2\bar\zeta_{\dot \g}
x^{\dot\g \g}\eta_{\g}+
\r^2\eta^2\right)+
 2(x^2+\r^2)\bar\zeta_{\dot a}\eta_{\b}\Big]
\cr &
(\bar{\psi}^{(1)}_{q})^{\dot\a}_{\dot a}=g^{-1/2}{4\sqrt 2 q_{\a}^{\dagger} C^{\a\b}\rho^2 
\over (x^2+\rho^2)^{5/2}}\Big[x_{\b}^{\dot\a}\bar\xi_{\dot a}-\d_{\dot a}^{\dot\a}
(x_{\b\dot\b}\bar\zeta^{\dot\b}+\rho^2\eta_{\b})\Big]
.
}}

With the latter result one can compute all the currents 
in the second equation of \firstorder\
and finally obtain the first order correction in $C$ to 
the antifundamental scalar $A^{\dagger (1)}$ 
\eqn\vvqws{\eqalign{
A^{\dagger (1)}_{\dot a}= &
{2 \r^2 q^{\dagger}_{\a} C^{\a\b}\over (x^2+\r^2)^{5/2}}
\Big[x_{\b\dot a}\left(\bar\zeta^2-2\bar\zeta_{\dot \g}
x^{\dot\g \g}\eta_{\g}+
\r^2\eta^2\right)+
 2(x^2+\r^2)\bar\zeta_{\dot a}\eta_{\b}\Big]
.
}}
Some remarks are now in order. From the solution (B.19) in the Appendix, 
one notices that cancellations at $h=1=k$ imply that $A^{\dagger (1)}$ behaves as $1/x^3$ at
infinity. Therefore it will not contribute to the effective action as we
will explain in section 7.
Moreover, the  part proportional to ${\cal K}$ vanishes at $h=1=k$.
Finally, we observe that the solutions for $A^{(1)}$ and $A^{\dagger (1)}$ are conjugate. This is not obvious a
priori but follows from the equations of motion after noticing that the ${\cal K}$
part of $A^{\dagger (1)}$ vanishes and that the prepotential for $\bar\psi_q^{(1)}$ is
$-i C^{\alpha\beta} \psi_{\beta}^{(0)}$ (after replacing $q \rightarrow q^{\dagger}$).

Finally the solution for the spinor $\psi^{(1)}$ is
\eqn\psuno{\eqalign{ \psi^{(1) \dot a}_{\a} & =
g^{1/2}{4 \sqrt 2 i \r^2 q_{\b}  C^{\;\;\b}_{\a}  
\over (x^2+\rho^2)^{5/2}} \left(
\bar\zeta^2 x^{\dot a}_{\g}\eta^{\g}-\r^2\eta^2\bar\zeta^{\dot a}
\right) \cr & = g^{1/2}{4 \sqrt 2 i \r^2 q_{\b}  C^{\;\;\b}_{\a}  
\over (x^2+\rho^2)^{5/2}}\left(\bar\zeta^2-\r^2\eta^2\right)\bar\xi^{\dot a}.
}}
As in the zero-th order case, this expression is 
obtained by acting with a covariant derivative on the equation and inverting the 
resulting Laplace operator.


\subsec{Second order in $C$} Finally, we complete the iteration solving 
the equations at second order in the deformation. They read
\eqn\secondorder{\eqalign{ & \bar{\cal D}^{\dot\a \a(0)}
{\cal D}_{\a \dot\a}^{(0)} A^{(2)} + ig k A^{\dot\a \a(1)} {\cal
D}^{(0)}_{\a\dot\a}A^{(1)}-{g^2\over 4} k^2 A^{\dot\a \a(1)}A^{(1)}_{\a\dot\a}
A^{(0)} =\cr & 
\hskip 4cm 
=h \sqrt 2 C^{\a\b}\Big[{\cal D}_{\a\dot\a}^{(0)}(\bar\l^{\dot\a}\psi_{\b}^{(1)})+{ig\over 2}
k A^{(1)}_{\a\dot\a}
(\bar\l^{\dot\a}\psi^{(0)}_{\b})\Big]
\cr &\bar{\cal D}^{\dot\a \a(0)}
{\cal D}_{\a \dot\a}^{(0)} A^{\dagger (2)} -ig k A^{\dot\a\a(1)} {\cal D}_{\a\dot\a}^{(0)}
A^{\dagger (1)}-{g^2\over 4} k^2 A^{\dot\a \a(1)}A^{(1)}_{\a\dot\a}A^{\dagger(0)} =
-i\sqrt 2 g \bar{\psi}_{\dot\a}^{(2)}\bar\l^{\dot\a} \cr &
\bar{{\cal D}}^{\dot\a \a (0)}\psi_{\a}^{(2)}+{ig\over 2} k A^{\dot\a \a (1)}\psi_{\a}^{(1)}=
-{\sqrt 2\over 2} g \bar\l^{\dot\a}A^{(2)}
\cr &
{\cal D}_{\a\dot\a}^{(0)}\bar{\psi}^{\dot\a (2)}-{ig \over 2} k A_{\a\dot\a}^{(1)}
\bar{\psi}^{\dot\a (1)}
= i h {\sqrt 2\over 2} C_{\a}^{\; \b}\Big[({\cal D}^{(0)}_{\b\dot\b}A^{\dagger (1)})\bar\l^{\dot\b}
-{ig\over 2} k A_{\b\dot\b}^{(1)}A^{\dagger (0)}\bar\l^{\dot \b}\Big].}} 

At this stage many simplifications take place as one can easily check by Grassmann counting. 
For example, the third equation drops out because $A^{\dot\a\a (1)}\psi^{(1)}_{\a}$ and $\bar\l^{\dot\a}A^{(2)}$ are of order 5 in $\bar\zeta$ and $\eta$ and therefore vanish. One would then be left  with the homogeneous equation which has no solution in the given instanton background.

Using the same procedure outlined in the previous sections one can solve \secondorder\ completely.
Details of the computations are contained in the Appendix B.

The solution for $A^{(2)}$ is quartic in the 
supersymmetry collective coordinates and reads
\eqn\aduemml{
A^{(2)\dot a}={C^2 q^{\a}x_{\a}^{\dot a} \bar\zeta^2\eta^2\over (x^2+\r^2)^{5/2}}
\left(\r^2-6x^2
\right)
}
where we have defined $C^2=C^{\a\b}C_{\a\b}=2 ~ {\rm det} ~ C$.

Using the usual Ansatz \psiansatza\ one can solve the fourth equation for $\bar\psi^{(2)}$. 
It consists of two contributions
\eqn\cdghdg{\bar\psi^{(2)}=\bar\psi^{(2)}_{{\cal K}}
+\bar\psi^{(2)}_{q}}
where
\eqn\barpsiduetext{\eqalign{
(\bar\psi^{(2)}_{{\cal K}})_{\dot\a \dot a}= &
g^{-1/2}{4{\cal K} C^2 \bar\zeta^2\eta^2 \e_{\dot\a \dot a}\over 3
(x^2+\r^2)^{7/2}}\left(
x^4+2x^2\r^2-5\r^4
\right) \cr 
(\bar\psi^{(2)}_{q})_{\dot\a \dot a}
= & g^{-1/2}{2\sqrt 2 h q^{\dagger}_{\g} C^2 \e_{\dot\a\dot a}
\over (x^2+\r^2)^{7/2}}\Big[
\bar\zeta^2\eta^{\g} (\r^4+2x^2\r^2-x^4)+2\r^4\eta^2 \bar\zeta^{\dot\g} x_{\dot\g}^{\g}
\Big]
.
}}

Finally, the solution for $A^{\dagger (2)}$ is
\eqn\bbbbnm{
A^{\dagger (2)}_{\dot a}=
-{C^2 q^{\dagger}_{\a}x^{\a}_{\dot a} \bar\zeta^2\eta^2\over (x^2+\r^2)^{5/2}}
\left(2\r^2+6x^2
\right).}
This solution goes as $1/x^2$ and  therefore will
contribute to the effective action presented in section 7.
The part of the solution proportional to ${\cal K}$ drops out because it contains too many powers of $\bar\zeta$ and $\eta$.

It is easy to check that there are no contributions of order  $C^3$ 
because of the Grassmannian nature of the collective coordinates which enter 
in the equations.


\newsec{The effective action}
At this point one can substitute the solutions to the equations 
of motion into the classical action and keep 
the leading terms. This gives the leading order 
correction in the coupling constant to the classical gauge instanton action. One can write
\eqn\ac{
S_{eff}={8\pi^2\over g^2}+g^0 S^{~ tot}_0}
where $S^{~ tot}_0$ is the sum of the contributions from untilded fields, $S_0$, and from tilded fields, $\tilde{S}_0$. 
Upon partial
integration and use of the equations of motion \eom, one has \eqn\ledact{ 
S_{0}= {1\over 2} \int  d^4x ~ \del^{\dot\a \a} \left( \left({\cal D}_{\a\dot\a} A^{\dagger}\right) A
\right)
}
and a similar expression for $\tilde{S}_0$.

Using Gauss's theorem, \ledact\ can be converted into a surface integral on the sphere at infinity. Then, one needs to keep only terms of order $1/x^3$.  
Analyzing the asymptotics of the different fields one is left with the following
\eqn\surftermexpl{\eqalign{
S_{0} ~ = ~ {1\over 2} {\rm vol}(S^3) ~ {x^{\dot\a \a}\over |x|} 
\left({\cal D}_{\a\dot\a}^{(0)}\left( A^{\dagger (0)} + A^{\dagger (2)}\right)\right)_{\dot a}
A^{(0)\dot a}\Big|_{|x|\rightarrow \infty}
.}}
Note that there are no contributions either from $A^{\dagger (1)}$ or the $U(1)$ part of the covariant derivative. As a consequence the effective action does not contain terms linear in the deformation parameter $C$ and is Lorentz invariant
\foot{In fact the final result will only depend on the scalar $C^2=2~{\rm det}~C$. This seems to be a general feature. For example, in \HatanakaRG\ it was shown that the non-anticommutative deformation of the chiral effective superpotential is also Lorentz invariant, even though the microscopic lagrangian is not.}.
After plugging in the explicit solutions 
\surftermexpl\ becomes
\eqn\iop{
S_{0}~ = ~ 2 \pi^2 |q|^2 \r^2 + 4 \pi^2 \sqrt 2 \r^2 q^{\a} \eta_{\a} {\cal K} + 24 \pi^2 |q|^2  C^2 \bar\zeta^2\eta^2
.}
Analogously, $\tilde{S}_0$ is
\eqn\ioptilde{
\tilde{S}_{0}~ = ~ 2 \pi^2 |\tilde{q}|^2 \r^2 + 4 \pi^2 \sqrt 2 \r^2  \tilde{{\cal K}}
\eta^{\a} \tilde{q}_{\a} + 24 \pi^2 |\tilde{q}|^2  C^2 \bar\zeta^2\eta^2
.}
Using the D-term constraint \dflasol\ one has $q^{\a}\eta_{\a}=q\eta_{1}$ and $\eta^{\a}\tilde{q}_{\a}=-\tilde{q}\eta_{2}$. One can also identify ${\cal K}=q^{\dagger}_{\a}{\cal K}^{\a}=q^{\dagger}{\cal K}^1$ and $\tilde{{\cal K}}={\cal K}_{\a}\tilde{q}^{\dagger \a}=\tilde{q}^{\dagger}{\cal K}^2$.
This, along with the condition $|q|^2=|\tilde{q}|^2$,  
allows one to write the total effective action in the form 
\eqn\toteffact{
S^{~ tot}_0=4\pi^2|q|^2\r^2+4\pi^2\sqrt 2 |q|^2 \r^2 \eta_{\a}{\cal K}^{\a} 
+48 \pi^2 |q|^2 C^2 \bar\zeta^2\eta^2.
}
We finally notice that, defining a "deformed instanton size" 
\eqn\defsize{\r^2_C \equiv \r^2+12C^2\bar\zeta^2\eta^2,}
the action \toteffact\ can be recast in a way which is formally equivalent to the undeformed case 
\eqn\toteffactundef{\eqalign{
S^{~ tot}_0= & 4\pi^2|q|^2\r^2_C+4\pi^2\sqrt 2 |q|^2 \r^2_C \eta_{\a}{\cal K}^{\a}\cr  \equiv & 4\pi^2 |q|^2 (\r^{inv}_{C})^2
}}
where $(\r^{inv}_{C})^2\equiv \r^2_{C}(1+\sqrt 2 \eta_{\a}{\cal K}^{\a})$. This combination will prove to be invariant under the ${\cal N}=1/2$ transformations of the moduli we present in the next section.



\newsec{Supersymmetry transformations of the moduli}
In order to check the supersymmetry invariance of the super-instanton effective
action at leading order \toteffact\ we need to know how the different  (pseudo)-collective
coordinates transform under ${\cal N}=1/2$ supersymmetry. To do this we only
need to look at the undeformed part of the solution, and equate active ${\cal
N}=1/2$ supersymmetry transformations acting on the fields to passive
transformations acting on the supersymmetric collective coordinates. In this
way performing an ${\cal N}=1/2$ supersymmetry variation on the background
yields another background of the same functional form, but with shifted
collective coordinates. This is the usual procedure, and we find
\eqn\shiftcollective{\eqalign{
\delta_\e (x_0)_{\alpha\dot\alpha}&=4i \e_\a \bar\zeta_{\dot\a}  \cr
\delta_\e \rho^2&=4i\rho^2\e^\a\eta_\a \cr
\delta_\e \bar\zeta^{\dot\a} &= 0 \cr
\delta_\e \eta_\a &=4i\eta_\a \e^\b \eta_\b \cr
\delta_\e {\cal K}_\a &=2\sqrt 2 i \e_\a -8i\eta^\b{\cal K}_\b\e_\a~.}}
which are the known transformations of the moduli for ${\cal N}=1$
supersymmetric Yang-Mills coupled to matter. 

It is straightforward to check
that the effective super-instanton action \toteffact\ is invariant under these
transformations. This occurs because of the delicate interplay between the $U(1)$ part of the connection (proportional to the parameter $k$ introduced before) and the Yukawa-like term (proportional to $h$). In fact, if one keeps $h$ and $k$ arbitrary one would get the following additional contributions to the effective action
\eqn\seffextra{S_{0}^{extra}=4 \pi^2 (h-k) q^{\dagger}_{\a}C^{\a}_{\;\;\b}q^{\b}(\bar\zeta^2-\r^2\eta^2)+4 \sqrt 2 \pi^2 (h-k){\cal K}\bar\zeta^2\eta_{\a}C^{\a}_{\;\;\b}q^{\b}}
with a similar contribution for the tilded fields.
We stress that the second term in \seffextra\ explicitly breaks the supersymmetry. 

                                                                                
Moreover, the usual transformations of the collective coordinates under the
antichiral supersymmetries are
\eqn\shiftcollectivebis{\eqalign{
\delta_{\bar\e} (x_0)_{\alpha\dot\alpha}&=0  \cr
\delta_{\bar\e} \rho^2&=0 \cr
\delta_{\bar\e} \bar\zeta^{\dot\a} &= \bar\e^{\dot\a} \cr
\delta_{\bar\e} \eta &=0\cr
\delta_{\bar\e} {\cal K}_\a &=0.}}
We see that the first two terms in \toteffact, which come from ${\cal N}=1$
supersymmetric Yang-Mills are invariant under \shiftcollectivebis, but the
third one is not. This is at it should, and the explicit breaking of ${\cal
N}=1$ to ${\cal N}=1/2$ is reflected at the level of the super-instanton
effective action.


\newsec{Gluino condensate}
In this final section we comment on how the superspace deformation
affects the gluino condensate. We begin by reviewing the
computation of the condensate in the instanton background with matter in the undeformed case. 
This method is usually referred to
as the weak coupling approach \AffleckMK \AffleckXZ \NovikovIC \DoreyIK.  

The fundamental idea is that adding matter
to pure gluodynamics, and assuming $\langle A \rangle\gg
\Lambda_{\rm {QCD}}$ allows one to do the calculation consistently in
the weakly coupled region. This way we prevent the running of the
gauge coupling into the strong coupling region where
semi-classical instanton techniques are not reliable. One can then
eventually decouple matter and go back to pure gluodynamics using
standard renormalization group and holomorphy arguments.  In the
weak coupling  approach we find the gluino condensate directly by
integrating the gauge invariant operator ${\rm Tr}\bar\l\bar\l$
over the instanton moduli space ${\cal M}$ \eqn\gluino{\int_{\cal
M} d\m\, {\rm Tr} \bar\l\bar\l(x)} where $d\m$ is the  instanton
measure \ShifmanMV \eqn\meas{d\m \sim {M_{\rm{{PV}}}^5\over q^2
}\left({8\pi^2\over{g^2}}\right)^2 e^{-S_{eff}} \; {d\rho^2 \over
\rho^2}\; d^4x_0 \; d^2\bar\zeta \; d^2\eta \; d^2 \cal{K}}
where $M_{\rm{{PV}}}$ is the Pauli-Villars regularization mass. Note in
particular the Grassmann integration over the fundamental
collective coordinates.
  Since the effective action does not
contain the $\bar\zeta$ collective coordinate corresponding to the
broken susy generator $\bar Q$, the only way to saturate the
$d^2\bar\zeta$ part of the measure is to pick the $\bar{\zeta}^2$
component of the gluino insertion. The Grassmann integral over the
remaining collective coordinates is saturated by pulling down
powers of the Yukawa term $\eta^{\a} \cal{K}_{\a}$ in the 't Hooft
effective action $S_{eff}$. The final result does not depend on
the location of the insertion. This follows directly also from the
fact $\bar\l\bar\l$ is the lowest component of an anti-chiral
superfield.


Now we turn to the ${\cal N}=1/2$ case. 
The measure \meas\ is altered by the $C$-dependent contributions coming from \foot{We thank Arkady Vainshtein for suggesting us to investigate this point.}
\eqn\defmeas{
\eqalign{ 
\left|\left|{\partial \bar\psi\over \partial \cal{K}}\right|\right|
\left|\left|{\partial \bar{\tilde{\psi}}\over \partial \tilde{\cal{K}}}\right|\right|\equiv &
\left(\int d^4x \left|{\partial \bar\psi\over \partial \cal{K}^{\a}}\right|^2\right)^{{1\over 2}}
\left(\int d^4x \left|{\partial \bar{\tilde{\psi}}\over \partial \cal{K}^{\a}}\right|^2\right)^{{1\over 2}}=\cr 
= & \pi^2 |q|^2 \left(\r^2+{4\over 3}(3h^2-k^2)C^2\bar\zeta^2\eta^2\right).}} 
For $h=1=k$ this is \eqn\defmeasbis{\left|\left|{\partial \bar\psi\over \partial \cal{K}}
\right|\right|
\left|\left|{\partial \bar{\tilde{\psi}}\over \partial \tilde{\cal{K}}}\right|\right|=\pi^2 |q|^2\left(\r^2+{8\over 3}C^2\bar\zeta^2\eta^2\right).}
This enters in the denominator of the measure \ShifmanMV.
The only contributions to $\left|\left|{\partial \bar\psi\over \partial \cal{K}}\right|\right|$ come from $\left|\left|{\partial \bar\psi^{(0)}\over \partial \cal{K}}
\right|\right|$ 
and 
$\left|\left|{\partial \bar\psi^{(1)}\over \partial \cal{K}}
\right|\right|$. 
It turns out that $\bar\psi^{(0)}$, $\bar\psi^{(1)}$, and $\bar\psi^{(2)}$ are mutually orthogonal. The undeformed part $\left|\left|{\partial \bar\psi^{(0)}\over \partial \cal{K}}
\right|\right|\left|\left|{\partial \bar{\tilde{\psi}}^{(0)}\over \partial \cal{K}}
\right|\right|$ gives the usual term $\pi^2 |q|^2 \r^2$.
From \defmeasbis\ and \defsize\ one notices that the deformation does not convert ${d\r^2\over \r^2}$ into 
${d\r_{C}^2\over \r_{C}^2}$, as one might have expected. 

As it is well-known, in the absence of the deformation, the integration of \meas\ (upon substitution of $q$ with a scalar superfield $\Phi$) yields the Affleck-Dine-Seiberg superpotential \AffleckMK. After turning on the deformation,  one can see that the additional contributions to the effective action and the measure do not modify the superpotential. Indeed, integrating over the fermionic coordinates $\cal{K}$ and $\eta$ first, the $C$-dependent pieces  vanish.

Even though the presence of the
deformation modifies the effective action and the measure, it is easy to verify
that \eqn\gluino{\int_{\cal M} d\m_{C}\,
e^{-48 \pi^2 |q|^2 C^2{\bar\zeta}^2{\eta}^2}{\rm Tr} \bar\l\bar\l(x)} does not
depend on $C$. Indeed the only way to saturate the integral over
the fundamental fermionic collective coordinate ${\cal K}^{\a}$ is
through the term ${\cal K}^{\a}\eta_{\a}$ in the undeformed effective
action $S_{eff}^{C=0}$. Doing so we saturate also the $d^2\eta$
associated to the broken superconformal transformations moduli.
There is no space left then for $C^2\bar\zeta^2\eta^2$. We conclude
that the gluino condensate is not modified.

We can support the above conclusion also through the following
formal argument.  It was noted in \ImaanpurIG\ that varying the
pure SYM Lagrangian with respect to the deformation parameter $C$
yields a $Q_{\a}$ exact term 
 \eqn\Cvar{\eqalign{ & \delta_{C} {\cal L}_{gauge} 
= \delta C^{\m\n}
 \left(iF_{\m\n}\bar\l\bar\l-{C_{\m\n}\over
4}(\bar\l\bar\l)^2\right)
 \sim \d C^{\m\n}\{Q^{\a},(\s_{\m\n})_{\a\b}\l^{\b}\bar\l
\bar\l\}.
}} 
The new thing in the present context is the $C$
variation of the matter sector. After setting the auxiliary fields to zero
this reads 
\eqn\iman{\delta_{C} {\cal
L}_{mat}=-{\sqrt 2\over 2}\delta{C}^{\a\b} {\cal
D}_{\alpha\dot\alpha} A^\dagger
\bar\lambda^{\dot\alpha}\psi_{\beta}-{\sqrt 2\over
2}\delta{C}^{\a\b}{\tilde \psi_{\beta}}
 \bar\lambda^{\dot\alpha}{\cal D}_{\alpha\dot\alpha}
{\tilde A}^\dagger ~ \sim ~ \d C^{\a\b}\{Q_{\a},\bar\psi\bar\l\psi_{\b}+\tilde\psi_{\b}
\bar\l
{\bar{\tilde
\psi}}\} .} 
Therefore the matter sector is also Q-exact
under the $C$-deformation.  Let us now apply this argument to
 discuss the dependence of the gluino condensate $\langle
\bar\l\bar\l\rangle$ on the
deformation parameter. The $C$ variation of the action inserts a
$\d_{Q_{\a}}(~\cdot ~)$ in the correlation function. The
$\delta_{Q_{\a}}$ operator can be pulled past the insertion to act
on the vacuum. Therefore we conclude that 
\eqn\fin{{\delta\over \delta
C_{\m\n}}\langle \bar\l \bar\l(x)\rangle=0} and we recover the
familiar ${\cal N}=1$ result.   We cannot make similar
considerations for antichiral insertions of $A^{\dagger} $ because
the profile now depends explicitly on $C$.  The $C$ deformation can give non trivial contributions to correlation functions like $\langle   A^\dagger \tilde{A}^\dagger\rangle$ and $\langle \bar\psi\bar{\tilde\psi} \rangle$.


\newsec{Conclusion}

In this paper we analyzed the effect of the ${\cal N}=1/2$ superspace deformation on the instanton calculus in presence of a matter superfield. The field equations  have been solved  iteratively in the deformation parameter and at leading order in the coupling constant. The fermionic nature of the
back-reaction allows to find the exact $C$ dependence of the solution.  The matter part receives linear and quadratic corrections in $C$. The asymptotic behavior of the $C$-deformed matter solution modifies the 't Hooft supersymmetric effective action. The correction is quadratic in $C$, quartic in the collective coordinates and ${\cal N}=1/2$ supersymmetric. The modified effective action does not alter the gluino condensate.


\vskip .5cm
{\bf Acknowledgements } It is a pleasure to 
thank Brenno Carlini Vallilo and Martin Ro\v cek 
 for useful discussions and Arkady Vainshtein
for helpful correspondence. 
We acknowledge partial financial support
through NSF award PHY-0354776. Opinions, findings, 
and conclusions expressed here are those of the authors 
and do not necessarily reflect the views of the 
National Science Fundation.


\appendix{A}{Notation and conventions}

In this appendix we define our notation and 
conventions and collect some useful formulas. 
 
{\bf Spinor algebra}
We work in Euclidean space $R^4$ but we continue to adopt
 the Lorenztian signature notation. The spinor notation is based 
on the $SU(2)_L \times SU(2)_R$ algebra of the Lorentz group. In Euclidean space 
the $SU(2)$ subalgebras are not related by complex conjugation. 

The spinor indices $\a,\dot\a=1,2$ are raised and lowered 
with the $\epsilon$ tensor in the following way
\eqn\aadfg{\eqalign{ &
\psiu = \epsu \psi _{\beta} \quad \qquad \psid = \epsd \psi ^{\beta}
\quad \cr &
\psiudot = \epsudot \barpsi _{\betadot} \quad 
\qquad \psiddot = \epsddot \barpsi ^{\betadot}.}}
Useful identities involving $\epsilon$ tensors are
\eqn\uuu{\eqalign{ &
\epsd \epsilon ^{\beta\gamma} = \delta _{\alpha}^{\gamma}
\quad ~~~~~~~~~~~~~~~~~~~
\epsddot \epsilon ^{\betadot\gammadot}
= \delta _{\alphadot}^{\gammadot} \quad \cr &
\epsd \epsilon ^{\delta\gamma} =
\delta _{\alpha}^{\gamma}\delta _{\beta}^{\delta} -
\delta _{\alpha}^{\delta}\delta _{\beta}^{\gamma}
\quad  ~~~~~~~
\epsddot \epsilon ^{\deltadot\gammadot} =
\delta _{\alphadot}^{\gammadot}\delta _{\betadot}^{\deltadot} -
\delta _{\alphadot}^{\deltadot}\delta _{\betadot}^{\gammadot}.}}
The spinors are contracted according to the following rules
\eqn\ii{\eqalign{
& \psi\chi = \psiu\chi_{\alpha} = -\psid\chi^{\alpha} =
\chi^{\alpha}\psid = \chi\psi \cr &
\barpsi\barchi = \psiddot\barchi^{\alphadot}
= -\psiudot\barchi_{\alphadot} =
\barchi_{\alphadot}\psiudot = \barchi\barpsi
}}
and
\eqn\pplo{\eqalign{ &
\psiu \psi ^{\beta} = -\half\epsu \psi\psi \quad ~~~~~~~~~
\psid \psi _{\beta} = \half\epsd \psi\psi \quad \cr &
\psiudot \barpsi ^{\betadot} = \half\epsudot \barpsi\barpsi
\quad ~~~~~~~~~~~
\psiddot \barpsi _{\betadot} = -\half\epsddot \barpsi\barpsi.}}
We adopt the following definition 
\eqn\equis{
x^2\equiv {1\over 2}x^{\dot{\alpha}\alpha}x_{\alpha\dot{\alpha}}=-x^{\m}x_{\m}} 
from which follows that
\eqn\ghjhjm{
x_{\alpha\dot{\beta}}x^{\dot{\beta}a}=\delta_{\alpha}^a x^2.}
The derivative acts as
\eqn\equisa
{
\partial_{\alpha\dot{\alpha}}x^{\beta\dot{\beta}}
=-2\delta_{\alpha}^{\beta}\delta_{\dot{\alpha}}^{\dot{\beta}}.
}

{\bf Covariant derivative}
The covariant derivative with the undeformed connection is 
\eqn\covderfund{
{\cal D}^{(0)\dot a}_{\a\dot\a \;\; \dot b}=\partial_{\a\dot\a}\d_{\dot b}^{\dot a}
\pm {1\over x^2+\rho^2}(\d_{\dot \a}^{\dot a}x_{\a\dot b}+\e_{\dot b\dot\a }x_{\a}^{\dot a})
}
The $+$ sign is used when the derivative acts on fundamental fields and the $-$ sign when it acts on  antifundamental ones.

{\bf Propagator }
In order to solve Poisson equations in the $SU(2)$ self-dual instanton 
background we repeatedly use the following propagator which is the inverse of $-({\cal D}^{(0)})^2$ \AmatiFT
\eqn\propagator{G^{\dot a}_{\; \dot b}(x,y)={i \over 4\pi^2}
{x^{\dot a\g}y_{\g\dot b} + \r^2\delta^{\dot a }_{\dot b}\over 
(x^2+\r^2)^{1/2}(y^2+\r^2)^{1/2}(x-y)^2}.}


\appendix{B}{Solutions order by order in $C$}

In this appendix we give details about the procedure for getting the solutions to the matter field equations order by order in $C$. In particular we present explicit expressions for the currents appearing in the Poisson equations.

\subsec{Zero-th order in $C$} 
The solution for $A^{\dagger (0)}$ is given by solving 
\eqn\adaggeq{
\left({\cal D}^{(0) 2}A^{\dagger (0)}\right)_{\dot a} 
= i{\sqrt 2 \over 2} g
\bar\psi^{(0)}_{\dot\a \dot b}\bar\l^{\dot \a \dot b}_{\;\;\;
\dot a}
=-{12\sqrt 2 {\cal K}\r^4\over (x^2+\r^2)^{7/2}}
\bar\xi_{\dot a} \equiv {\cal J}_{(A^{\dagger})\dot a} .}
The operator $({\cal D}^{(0)})^{2}$ can be inverted using \propagator. The convolution
\eqn\convoll{A^{\dagger (0)}_{\dot a}(x)=-\int d^4y {\cal J}_{(A^{\dagger})\dot b}(y) G^{\dot b}_{\; \dot a}(y,x)} 
yields a particular solution. Adding the homogeneuos solution (which is, up to the position of the indices, the same as for $A^{(0)}$) one obtains
\eqn\adagge{A^{\dagger (0)}_{\dot a}(x,\r,{\cal K},\bar\xi)=q^{\dagger}_{\a}{x^{\a}_{\dot a}\over 
\sqrt{x^2+\r^2}}-{\sqrt 2 {\cal K} \r^2 \bar\xi_{\dot a}\over (x^2+\r^2)^{3/2}}.}

The solution for $\psi^{(0)}$ is
given by acting with a covariant derivative on both sides of its equation in \zerothorder\ and by solving  \eqn\psiii{\left({\cal D}^{(0) 2}\psi_{\a}^{(0)}\right)^{\dot a}={\sqrt 2 \over 2} g 
{\cal D}_{\a\dot\a}^{(0)}
(\bar\l^{\dot\a \dot a}_{\;\;\;\;\dot b} A^{(0)\dot b})\equiv 
{\cal J}_{(\psi)\a}^{\dot a}
} where the current is \eqn\cur{\eqalign{{\cal J}^{\a \dot a}_{(\psi)}& = g^{1/2} {4 \sqrt 2
 i q^{\b} \r^2 \over
(x^2+\r^2)^{7/2}}\Big[\d^{\a}_{\b}\bar\xi^{\dot a}(x^2+6\r^2)+
x_{\b}^{\dot a}x_{\dot c}^{\a}\bar\xi^{\dot c}-x^{\dot a
\a}x_{\b \dot c}\bar\xi^{\dot c}\Big]\cr & =g^{1/2}{24\sqrt 2
 i q^{\a}\rho^4 
\over (x^2+\rho^2)^{7/2}}\bar\xi^{\dot a}.}}
The last line is obtained by using the following identity
\eqn\id{x^{\dot a}_\a x^{\dot b}_\b-x^{\dot b}_\a x^{\dot a}_\b=-\epsilon_{\a\b}
\epsilon^{\dot a \dot b}x^2.}
By convoluting with the propagator in \propagator\ one gets (in this case the homogeneous solution is vanishing)
\eqn\psiiii{\psi^{\a \dot a (0)}(x,\rho,\bar\xi)=
-g^{1/2}{2 \sqrt{2} i q^{\a}  \rho^2 \over (x^2+\rho^2)^{3/2}}\bar\xi^{\dot a}.}



\subsec{First order in $C$}

One starts by plugging in the first equation in \firstorder\ the explicit solutions for the fields at zero-th order in $C$. After taking all the derivatives one can rewrite the equation explicitly as  
\eqn\vv{\eqalign{ 
\left({\cal D}^{(0) 2} A^{(1)}\right)^{\dot a} 
= & -{16 k \r^2 q_{\a} C^{\a\b}\over (x^2+\r^2)^{9/2}}\Big[x_{\b}^{\dot a}(x^2+3\r^2) \bar\zeta^2 - 2 x_{\b}^{\dot a} x^2 \r^2 \eta^2 -\cr & \hskip 2.5cm -
2 \r^2 \left(2 x_{\b}^{\dot a}  \bar\zeta_{\dot \g} x^{\dot \g \g}\eta_{\g}-(x^2+\r^2)\bar\zeta^{\dot a}\eta_{\b}\right)\Big]+ \cr &
+{48 h \r^4 q_{\a} C^{\a\b}\over (x^2+\r^2)^{9/2}}\Big[2 x_{\b}^{\dot a}\bar\zeta^2+x_{\b}^{\dot a}\eta^2 (\r^2-x^2)-\cr & \hskip 2.5cm  -2\left(2 x_{\b}^{\dot a}  \bar\zeta_{\dot \g} x^{\dot \g \g}\eta_{\g}-(x^2+\r^2)\bar\zeta^{\dot a}\eta_{\b}\right)\Big].
}}
Using $\bar{\cal D}^{\dot\a \a(0)}{\cal D}_{\a \dot\a}^{(0)}=-2({\cal D}^{(0)})^2$ and taking the convolution with \propagator\ one gets 
\eqn\fundscalaapp{\eqalign{
A^{(1) \dot a}
= & {q_{\a} C^{\a\b}\over (x^2+\r^2)^{5/2}}
\Big[x^{\dot a}_{\b}\left(2(h-k)x^2+(5h-3k)\r^2\right)\bar\zeta^2+
\cr & \hskip 2cm
+x^{\dot a}_{\b}\left(2(k-h)x^2+(h+k)
\r^2\right)\r^2\eta^2+
\cr &\hskip 2cm +
 2\r^2(k-3h)\left(x^{\dot a}_{\b}\bar\zeta_{\dot \g}
x^{\dot\g \g}\eta_{\g}-(x^2+\r^2)\bar\zeta^{\dot a}\eta_{\b}\right)\Big].
}}
By putting $k=1$ and $h=1$ this result becomes \fundscala.

The equation for $\bar\psi^{(1)}$ in \firstorder\ can be rewritten in a more manageable way 
\foot{Using the fact that the $U(1)$ part of the connection can be rewritten as \eqn\aax
{A_{\a\dot\a}^{(1)}= - C_{\a}^{\;\b}\del_{\b\dot\a}K=- C_{\a}^{\;\b}{\cal D}^{(0)}_{\b\dot\a}K} one could think to rewrite this equation
as a total covariant derivative
\eqn\ccchg{{\cal D}^{\b (0)}_{\dot\a}\Big[
\epsilon_{\a\b}\bar\psi^{\dot \a(1)}+{iC_{\a\b}\over 2}(K\bar\psi^{\dot\a  (0)}-2\sqrt 2 A^{\dagger (0)}\bar\l^{\bar\a})\Big]=0} and then obtain a solution by  putting to zero the argument of the derivative. Actually, this does not yield a solution since $C_{\a\b}$ is a symmetric tensor.} 
using the Ansatz already discussed in the main text
\eqn\psia{\bar\psi^{\dot \a (1)}=\bar{\cal D}^{\dot \a \b}\Psi^{(1)}_{\b}.}
The equation then becomes 
\eqn\oool{({\cal D}^{(0)})^2\Psi_{\a}^{(1)}
=-{ig\over 2} k A_{\a\dot\a}^{(1)}\bar\psi^{\dot\a (0)}-i h {\sqrt 2\over 2} C_{\a}^{\;\b}({\cal D}_{\b\dot\b}^{(0)}A^{\dagger (0)})\bar\l^{\dot \b}\equiv {\cal J}_{\a}^{(\bar\psi^{(1)})}.} 
The explicit expression for the current reads
\eqn\ghhu{\eqalign{
{\cal J}_{\a\dot a}^{(\bar\psi^{(1)})}= &
-{8 k \r^2 C_{\a}^{\;\;\b}{\cal K}\over (x^2+\r^2)^{9/2}}\Big[
x_{\b \dot a}(x^2+3\r^2) \bar\zeta^2 - 2 x_{\b \dot a} x^2 \r^2 \eta^2 -\cr & \hskip 2.5cm -
2 \r^2 \left(2 x_{\b \dot a}  \bar\zeta_{\dot \g} x^{\dot \g \g}\eta_{\g}-(x^2+\r^2)\bar\zeta_{\dot a}\eta_{\b}\right)
\Big]+ \cr & + {24 h \r^4 {\cal K} C_{\a}^{\;\;\b} \over (x^2+\r^2)^{9/2}}
\Big[2 x_{\b \dot a}\bar\zeta^2+x_{\b \dot a}\eta^2 (\r^2-x^2)-\cr & \hskip 2.5cm  -2\left(2 x_{\b \dot a}  \bar\zeta_{\dot \g} x^{\dot \g \g}\eta_{\g}-(x^2+\r^2)\bar\zeta_{\dot a}\eta_{\b}\right)\Big] -
\cr & 
-{24 \sqrt 2 h \r^4 q_{\b}^{\dagger}C_{\a}^{\;\;\b} \over (x^2+\r^2)^{7/2}}\bar\xi_{\dot a}
.}}
Taking the convolution of  \ghhu\ with \propagator\ one obtains the prepotential
\eqn\cfgvgj{\eqalign{
\bar\Psi_{\a\dot a}^{(1)}= &
{{\cal K} C_{\a}^{\;\;\b}\over 2(x^2+\r^2)^{5/2}}
\Big[x_{\b\dot a}\left(2(h-k)x^2+(5h-3k)\r^2\right)\bar\zeta^2+
\cr & \hskip 2cm
+x_{\b\dot a}\left(2(k-h)x^2+(h+k)
\r^2\right)\r^2\eta^2+
\cr &\hskip 2cm +
 2\r^2(k-3h)\left(x_{\b\dot a}\bar\zeta_{\dot \g}
x^{\dot\g \g}\eta_{\g}-(x^2+\r^2)\bar\zeta_{\dot a}\eta_{\b}\right)\Big]-\cr & -
{2\sqrt 2 h \r^2 q^{\dagger}_{\b} C_{\a}^{\;\;\b}\over (x^2+\r^2)^{3/2}}\bar\xi_{\dot a}
.
}}
Acting on this with a covariant derivative we finally obtain 
\eqn\psiunoapp{\bar\psi^{(1)}=\bar\psi^{(1)}_{\bar{\zeta}^2}
+\bar\psi^{(1)}_{\eta^2}+\bar\psi^{(1)}_{\bar{\zeta}\eta}+\bar\psi^{(1)}_{linear}}
where
\eqn\psiunosolutionapp{
\eqalign{ &
(\bar\psi^{(1)}_{\bar{\zeta}^2})^{\dot \a}_{\dot a}=g^{-1/2}
{2 {\cal K} C_{\a}^{\; \b} x^{\dot \a \a} x_{\b\dot a}\bar{\zeta}^2
\over (x^2+\rho^2)^{7/2}}\left((h-k)x^2+2(2h-k)\rho^2\right)
\cr & (\bar\psi^{(1)}_{\eta^2})^{\dot \a}_{\dot a}=
g^{-1/2}{2{\cal K} C_{\a}^{\;\b} x^{\dot\a \a} x_{\b \dot a}\rho^2\eta^2
\over (x^2+\rho^2)^{7/2}}\left((k-h)x^2+2h\rho^2\right)\cr &
(\bar\psi^{(1)}_{\bar{\zeta}\eta})^{\dot\a}_{\dot a}=
g^{-1/2}
{4{\cal K} C_{\a}^{\;\b}x^{\dot\a \a}\rho^2\over
(x^2+\rho^2)^{7/2}}(k-3h)\left(x_{\b\dot a} \bar{\zeta}_{\dot \g}
x^{\dot\g \g}\eta_{\g}-(x^2+\rho^2)\bar\zeta_{\dot a}\eta_{\b}\right)
\cr &
(\bar{\psi}^{(1)}_{linear})^{\dot\a}_{\dot a}=g^{-1/2}{4\sqrt 2 h q_{\a}^{\dagger} C^{\a\b}\rho^2 
\over (x^2+\rho^2)^{5/2}}\left(x_{\b}^{\dot\a}\bar\xi_{\dot a}-\d_{\dot a}^{\dot\a}
(x_{\b\dot\b}\bar\zeta^{\dot\b}+\rho^2\eta_{\b})\right)
.
}}
One has \psiuno\ by setting $k=1$ and $h=1$.

Having solved for $\bar\psi^{(1)}$ allows one to tackle the equation for $A^{\dagger (1)}$. Explicitly this equation reads
\eqn\vvq{\eqalign{
\left({\cal D}^{(0) 2} A^{\dagger (1)}\right)_{\dot a} 
= & 
-{16 k \rho^2 q^{\dagger}_{\a} C^{\a\b} \over (x^2+\rho^2)^{9/2}}\Big[
x_{\b \dot a}(x^2+3\r^2) \bar\zeta^2 - 2 x_{\b \dot a} x^2 \r^2 \eta^2 -\cr & \hskip 2.5cm -
2 \r^2 \left(2 x_{\b \dot a}  \bar\zeta_{\dot \g} x^{\dot \g \g}\eta_{\g}-(x^2+\r^2)\bar\zeta_{\dot a}\eta_{\b}\right)
\Big]-\cr & 
-{48 \sqrt 2 k \r^4 {\cal K} C^{\a\b} x_{\b\dot a} \over (x^2+\rho^2)^{9/2}}\left(
\bar\zeta^2 \eta_{\a}-x_{\a\dot\b}\eta^2\bar\zeta^{\dot\b}
\right)
+ \cr & +
{48 h \r^4 q_{\b}^{\dagger}C_{\a\b} \over (x^2+\r^2)^{9/2}}
\Big[2 x_{\b \dot a}\bar\zeta^2+x_{\b \dot a}\eta^2 (\r^2-x^2)-\cr & \hskip 2.5cm  -2\left(2 x_{\b \dot a}  \bar\zeta_{\dot \g} x^{\dot \g \g}\eta_{\g}-(x^2+\r^2)\bar\zeta_{\dot a}\eta_{\b}\right)\Big]
+ \cr & +
{8 \sqrt 2 \r^2 {\cal K} C^{\a\b} x_{\b\dot a}\over (x^2+\r^2)^{9/2}}
\Big[
\left( (h-k)x^2 + (9h-3k)\r^2 \right)\bar\zeta^2\eta_{\a}+
\cr & \hskip 2.5cm + (2k-8h)\r^2x_{\a\dot\b}\eta^2 \bar\zeta^{\dot\b}
\Big]
.
}}
The convolution with \propagator\ yields 
\eqn\vvqwsapp{\eqalign{
A^{\dagger (1)}_{\dot a}= &
{q^{\dagger}_{\a} C^{\a\b}\over (x^2+\r^2)^{5/2}}
\Big[x_{\b\dot a}\left(2(h-k)x^2+(5h-3k)\r^2\right)\bar\zeta^2+
\cr & \hskip 2cm
+x_{\b\dot a}\left(2(k-h)x^2+(h+k)
\r^2\right)\r^2\eta^2+
\cr &\hskip 2cm +
 2\r^2(k-3h)\left(x_{\b\dot a}\bar\zeta_{\dot \g}
x^{\dot\g \g}\eta_{\g}-(x^2+\r^2)\bar\zeta_{\dot a}\eta_{\b}\right)\Big]+
\cr & +
{2\sqrt 2 {\cal K} C^{\a\b}x_{\a\dot a}\over (x^2+\r^2)^{5/2}}(h-k)
\Big[(x^2+2\rho^2)\bar\zeta^2\eta_{\b}-\rho^2\eta^2\bar\zeta^{\dot\b}x_{\b\dot\b}
\Big].
}}
The final solution with $k=1=h$ is given in \vvqws.

The equation for $\psi^{(1)}$ can be cast into
\eqn\bnmdc{\eqalign{ &
\left({\cal D}^{(0) 2}\psi_{\a}^{(1)}\right)^{\dot a}= \cr & \hskip 0.5cm
={16i\sqrt 2 \r^2 q_{\g} C^{\g\b} \over (x^2+\r^2)^{9/2}} \bar\zeta^2\Big[
\e_{\a\b}x_{\l}^{\dot a}\eta^{\l}\left((h-3k)x^2+(9k-3h)\r^2 \right) +
\cr & \hskip 6cm + x_{\a}^{\dot a}\eta_{\b}(k-h)\left( x^2+9\r^2 \right)
\Big]+\cr & \hskip 0.5cm +
{8 i \sqrt 2 \r^4 q_{\g} C^{\g\b} \over (x^2+\r^2)^{9/2}} \eta^2 \Big[
\e_{\a\b}\bar\zeta^{\dot a}\left( (k+5h)x^2-(3h+7k)\r^2 \right)+16(h-k)
x_{\b}^{\dot a}x_{\a\dot\a}\bar\zeta^{\dot\a}
\Big].}}
Inverting the covariant Laplacian we get 
\eqn\psunoapp{\psi^{(1)}=\psi^{(1)}_{\eta^2\bar\zeta}+\psi^{(1)}_{\bar\zeta^2\eta}}
with
\eqn\psunosolutionapp{
\eqalign{ 
(\psi^{(1)}_{\eta^2\bar\zeta})^{\dot a}_{\a} &=g^{1/2}
{2 \sqrt 2 i q_{\g}  C^{\g\b} \rho^2\eta^2 
\over (x^2+\rho^2)^{5/2}}\Big[2(h-k)x^{\dot a}_{\b}x_{\a\dot\a}\bar\zeta^{\dot \a}
+\rho^2\e_{\b\a}\bar\zeta^{\dot a}(k+h)
\Big] \cr 
(\psi^{(1)}_{\bar\zeta^2\eta})^{\dot a}_{\a}& =
g^{1/2}{2\sqrt 2 i q_{\g}  C^{\g\b} \bar\zeta^2
\over (x^2+\rho^2)^{5/2}}
\Big[\rho^2(3k-h)\e_{\a\b}x^{\dot a}_{\l}\eta^{\l}+
2(k-h)x^{\dot a}_{\a}\eta_{\b}(x^2+2\rho^2)\Big]. 
}}
This yields \psuno\ when $k=1=h$.


\subsec{Second order in $C$}

The first equation of \secondorder\ reads explicitly 
\eqn\aduem{\eqalign{
\left({\cal D}^{(0) 2} A^{(2)}\right)^{\dot a}= &
{32 k^2 \r^4 C^2 q^{\a}x^{\dot a}_{\a}\bar\zeta^2 \eta^2\over (x^2+\r^2)^{9/2}}
-
{64 k^2 \r^2 C^2 q^{\a}x^{\dot a}_{\a}\bar\zeta^2 \eta^2\over (x^2+\r^2)^{9/2}}(2x^2-\r^2)
-\cr & -
{16 h \r^2 C^2 q^{\a}x^{\dot a}_{\a}\bar\zeta^2 \eta^2\over (x^2+\r^2)^{9/2}}
\left((h-k)x^2+3(h+k)\r^2\right)
+{96 h k \r^4 C^2 q^{\a}x^{\dot a}_{\a}\bar\zeta^2 \eta^2\over (x^2+\r^2)^{9/2}}
}}
where we have defined $C^2=C^{\a\b}C_{\a\b}$, and therefore $C_{\a\b}C^{\b\g}={1\over 2}\d_{\a}^{\g}C^2$.
The solution to this equation is 
\eqn\aduemmlapp{
A^{(2)\dot a}={C^2 q^{\a}x_{\a}^{\dot a} \bar\zeta^2\eta^2\over (x^2+\r^2)^{5/2}}
\left( (-6k^2+2hk-2h^2)x^2+(k^2+3hk-3h^2)\r^2
\right).
}
The final result with $k=1=h$ is reported in the main text \aduemml.

The equation for $\bar\psi^{(2)}$ is again easily rewritten using the Ansatz \psia.
In terms of the prepotential we then obtain
\eqn\gty{\eqalign{  
\left({\cal D}^{(0) 2} \bar\Psi^{(2)}_{\a}\right)_{\dot a}= & 
g^{-1/2}{16\sqrt 2 h \r^2 q^{\dagger}_{\g}  \over (x^2+\r^2)^{9/2}}\bar\zeta^2 \Big[
C^2\d_{\a}^{\g}x_{\d\dot a}\eta^{\d}\left((h-k)x^2+3h\r^2\right)-C^2\eta^{\g}kx_{\a\dot a}x^2
+ \cr &
\hskip 1cm +
C_{\a\b}C^{\g\d}\left(6 k \r^2 \eta_{\d} x_{\dot a}^{\b}-\eta^{\b}x_{\d\dot a}
\left( (h-k)x^2 +(9h-3k)\r^2\right)\right)\Big]
+\cr &
+g^{-1/2}{4\sqrt 2 h \r^4 q^{\dagger}_{\g} \over (x^2+\r^2)^{9/2}}\eta^2 \Big[
C^2\d_{\a}^{\g}\bar\zeta_{\dot a}\left((5h+5k)x^2+(k-3h)\r^2\right)+
\cr & \hskip 1cm 
+4 k 
C^2\bar\zeta_{\dot\g}
x_{\a\dot a}x^{\dot\g \g}
- 
8 C_{\a\b}C^{\g\d}\left(3 k  \bar\zeta^{\dot\g} x_{\d\dot\g}x^{\b}_{\dot a}
+\bar\zeta^{\dot \g}x_{\d\dot a}x^{\b}_{\dot\g}(k-4h)\right)\Big]+
\cr &
+g^{-1/2}{8 \r^2 {\cal K}  C^2 \bar\zeta^2\eta^2 x_{\a\dot a}
\over (x^2+\r^2)^{9/2}}\left((3k-h)(h-k)x^2+(3h^2+k^2)\r^2 \right).
}}
Inverting the Laplacian yields
\eqn\tyu{\eqalign{ 
\bar\Psi^{(2)\a}_{\dot a}= &
g^{-1/2}{\sqrt 2 h q^{\dagger}_{\g}  \over 2(x^2+\r^2)^{5/2}}\bar\zeta^2\Big[
2 C^2 \e^{\a\g} \eta^{\b}x_{\b\dot a}(h-k)\left(2x^2+3\r^2\right)
- \cr & \hskip 1cm -C^2 k \eta^{\g}x_{\dot a}^{\a}(2x^2+\r^2)+ 8C^{\a}_{\;\;\b}
C^{\g\d}x_{\d\dot a}\eta^{\b}(k-h)(x^2+2\r^2)
\Big]
+\cr &
+g^{-1/2}{\sqrt 2 h \r^2 q^{\dagger}_{\g} \over 2(x^2+\r^2)^{5/2}}\eta^2\Big[
\r^2 C^2 \e^{\a\g}\e_{\dot a \dot\g}\bar\zeta^{\dot\g}(k-2h)-\cr &
\hskip 1cm -
kC^2\bar\zeta^{\dot\g}x_{\dot\g}^{\a}x_{\dot a}^{\g}+C^{\a\b}C^{\g\d}\bar\zeta^{\dot\g}
(8k-8h)x_{\d\dot a}x_{\b\dot\g}
\Big]+\cr &
+g^{-1/2} {{\cal K} C^2  x_{\dot a}^{\a}\over 3
(x^2+\r^2)^{5/2}}\bar\zeta^2\eta^2\left( 2k(3h-2k)x^2 + (3h^2+3hk-k^2)\r^2\right).
}}
Finally, the covariant derivative on \tyu\ yields the result for $\bar\psi^{(2)}$. 
It consists of three contributions
\eqn\cdghdgapp{\bar\psi^{(2)}=\bar\psi^{(2)}_{\bar\zeta^2 \eta}
+\bar\psi^{(2)}_{\eta^2\bar\zeta}+\bar\psi^{(2)}_{\bar\zeta^2\eta^2}}
where
\eqn\barpsiduetextapp{\eqalign{
(\bar\psi^{(2)}_{\bar\zeta^2\eta})_{\dot\a \dot a}= &
g^{-1/2}{\sqrt 2 h q^{\dagger}_{\g} \bar\zeta^2 \over (x^2+\r^2)^{7/2}}\Big[
\eta^{\d}C^2 x_{\d \dot a}x_{\dot\a}^{\g}
(4h-4k)(x^2+2\r^2)-\cr & 
-\eta^{\g}C^2\e_{\dot\a \dot a}(2kx^4-4kx^2\r^2-2h\r^4)+
\eta^{\b} C^{\a}_{\;\;\b}C^{\g\d} x_{\a \dot\a}x_{\d\dot a}(8h-8k)(x^2+3\r^2)
\Big]\cr 
(\bar\psi^{(2)}_{\eta^2\bar\zeta})_{\dot\a \dot a}
= & g^{-1/2}{\sqrt 2 h q^{\dagger}_{\g} \r^2 \eta^2 \over (x^2+\r^2)^{7/2}}\Big[
\bar\zeta^{\dot\g} \r^2 C^2 x^{\g}_{\dot\g}\e_{\dot\a\dot a}
(6k-2h)+\cr & \hskip 2cm +\bar\zeta_{\dot a} \r^2 C^2 x_{\dot\a}^{\g}
(4k-4h)+\bar\zeta^{\dot\g} C^{\a\b}C^{\g\d} x_{\b\dot\g} x_{\a\dot\a}
x_{\d\dot a}(16h-16k)\Big]\cr
(\bar\psi^{(2)}_{\bar\zeta^2\eta^2})_{\dot\a \dot a}= &
g^{-1/2}{4{\cal K} C^2 \bar\zeta^2\eta^2 \e_{\dot\a \dot a}\over 3
(x^2+\r^2)^{7/2}}\Big[
(3hk-2k^2)x^4+\cr & \hskip 2cm +(3h^2+5k^2-6hk)x^2\r^2+(k^2-3h^2-3hk)\r^4
\Big].
}}
The final result for $k=1=h$ is given in \barpsiduetext.

The equation for $A^{\dagger (2)}$ reads
\eqn\bbbb{\eqalign{
\left({\cal D}^{(0) 2}
  A^{\dagger (2)}\right)_{\dot a}= &
{32 k^2 \r^4 C^2 q^{\dagger}_{\a}x^{\a}_{\dot a}\bar\zeta^2 \eta^2\over (x^2+\r^2)^{9/2}}
-
{64 k^2 \r^2 C^2 q^{\dagger}_{\a}x^{\a}_{\dot a}\bar\zeta^2 \eta^2\over (x^2+\r^2)^{9/2}}(2x^2-\r^2)-
\cr & -
{8 h \r^2 C^2 q^{\dagger}_{\a}x_{\dot a}^{\a}\bar\zeta^2 \eta^2\over (x^2+\r^2)^{9/2}}
\left((2h-5k)x^2+(6h+3k)\r^2\right)
.}}
The convolution with \propagator\ yields 
\eqn\bbbbnmapp{
A^{\dagger (2)}_{\dot a}=
{C^2 q^{\dagger}_{\a}x^{\a}_{\dot a} \bar\zeta^2\eta^2\over (x^2+\r^2)^{5/2}}
\left( (-6k^2+2hk-2h^2)x^2+(k^2-3h^2)\r^2
\right).}
In \bbbbnm\ we report the explicit solution with $k=1=h$.

The equation for $\psi^{(2)}$ drops out as already explained in the main text.


\appendix{C}{Table of integrals}

In solving our systems of coupled differential equations the following integrals turn out to be very useful
\eqn\tableintegrals{\eqalign{ &
{\cal I}_{(0)}^p\equiv\int d^4y {1\over (y^2+\r^2)^p(x-y)^2}={i\pi^2\over p-1}
\int _0^1dz {1\over (\r^2+zx^2)^{p-1}}\cr &
{\cal I}_{(1)\a\dot\a}^p\equiv\int d^4y{y_{\a\dot\a}\over (y^2+\r^2)^p(x-y)^2}={i\pi^2\over p-i}
\int_0^1 dz {zx_{\a\dot\a}\over (\r^2+zx^2)^{p-1}}\cr &
{\cal I}^p_{(2)\a\dot\a\b\dot\b}\equiv\int d^4y {y_{\a\dot\a}y_{\b\dot\b}\over (y^2+\r^2)^p(x-y)^2}
=\cr & \hskip 3cm =
{i\pi^2\over p-1}\int_0^1 dz \Big[{(1-z)\e_{\a\b}\e_{\dot\a\dot\b}\over (p-2)(\r^2+zx^2)^{p-2}}+
{z^2x_{\a\dot\a}x_{\b\dot\b}\over (\r^2+zx^2)^{p-1}} \Big]\cr &
{\cal I}^p_{(3)\a\dot\a\b\dot\b\g\dot\g}\equiv
\int d^4y {y_{\a\dot\a}y_{\b\dot\b}y_{\g\dot\g}\over (y^2+\r^2)^p(x-y)^2}=\cr & \hskip 3cm
={i\pi^2\over p-1}\int_0^1 dz \Big[{z(1-z)(\e_{\a\b}\e_{\dot\a\dot\b}x_{\g\dot\g}+perm.)
\over (p-2)(\r^2+zx^2)^{p-2}}+
{z^3x_{\a\dot\a}x_{\b\dot\b}x_{\g\dot\g}\over (\r^2+zx^2)^{p-1}} \Big].
}}
These are the prototypes for all other integrals used in this paper, which can be obtained 
using the following recursion formulae
\eqn\gyuxgi{\eqalign{ & {\cal I}^p_{(2k)}\equiv \int d^4y {(y^2)^k\over (y^2+\r^2)^p(x-y)^2}={\cal I}^{p-1}_{(2k-2)}-\r^2{\cal I}^{p}_{(2k-2)}\cr & 
{\cal I}^p_{(2k+1)\a\dot\a}\equiv \int d^4y {y_{\a\dot\a}(y^2)^k\over (y^2+\r^2)^p(x-y)^2}={\cal I}^{p-1}_{(2k-1)\a\dot\a}-\r^2{\cal I}^{p}_{(2k-1)\a\dot\a} \cr &
{\cal I}^p_{(2k+2)\a\dot\a\b\dot\b}\equiv \int d^4y {y_{\a\dot\a}
y_{\b\dot\b}(y^2)^k\over (y^2+\r^2)^p(x-y)^2}={\cal I}^{p-1}_{(2k)\a\dot\a\b\dot\b}-\r^2{\cal I}^{p}_{(2k)\a\dot\a\b\dot\b}. 
}}


\listrefs

\bye